\documentclass[conference]{IEEEtran}
\IEEEoverridecommandlockouts

\usepackage{tikz}
\usepackage{xcolor}

\newcommand{\circlednum}[1]{%
  \tikz[baseline=(char.base)]{\node[shape=circle, draw=red!80!black, fill=red!80!black, inner sep=0.3pt, text=white] (char) {#1};}%
}
\usepackage{booktabs}   
\usepackage{adjustbox}
\usepackage{amsmath,amsfonts}
\usepackage{algorithmic}
\usepackage[ruled, linesnumbered, norelsize]{algorithm2e}
\usepackage{graphicx}
\usepackage{textcomp}
\usepackage[dvipsnames]{xcolor}
\usepackage[table]{xcolor}
\definecolor{lightred}{HTML}{FEECEC} 
\usepackage{enumitem} 
\usepackage{pifont}   
\usepackage{tikz}

\usepackage[T1]{fontenc}
\usepackage{multirow}
\usepackage{textcomp}
\usepackage{listings}
\usepackage{xcolor}
\usepackage{enumitem}
\usepackage{xurl}   

\newcommand{\mycircle}[1]{%
    \tikz[baseline=(char.base)]{
        \node[shape=circle, fill=black, inner sep=0.3pt] (char) {\textcolor{white}{\textsf{\textbf{#1}}}};
    }%
}

\usepackage{geometry}
\usepackage{tikz}
    \usetikzlibrary{positioning}
    \usetikzlibrary{arrows}
    \usetikzlibrary{backgrounds}
    \usetikzlibrary{arrows.meta}
\usepackage{amsmath}

\usepackage[ruled, lines numbered, norelsize]{algorithm2e}
\usepackage{booktabs, multirow}
\usepackage{enumitem}
\usepackage{tabularx}
\usepackage{array}
\newcolumntype{M}[1]{>{\centering\arraybackslash}m{#1}}
\usepackage{graphicx}
\usepackage{sidecap}
\usepackage{xcolor}
\usepackage[table]{xcolor}
\definecolor{crimson}{rgb}{0.86, 0.08, 0.24}
\usepackage{subcaption}
\usepackage{cancel}
\usepackage{mathtools}
\definecolor{THEMECOLOR}{RGB}{118, 185, 0}

\newcommand{\thiswork}{ENEC}

\usepackage{subcaption} 

\usepackage{cite}
\usepackage[hidelinks]{hyperref}
\def\BibTeX{{\rm B\kern-.05em{\sc i\kern-.025em b}\kern-.08em
    T\kern-.1667em\lower.7ex\hbox{E}\kern-.125emX}}
\begin{document}

\pdfpagewidth=8.5in
\pdfpageheight=11in

\newcommand{\iscasubmissionnumber}{800}

\pagenumbering{arabic}

\title{ENEC: A Lossless AI Model Compression Method Enabling Fast Inference on Ascend NPUs}
 \author{

 \IEEEauthorblockN{
 Jinwu Yang\IEEEauthorrefmark{1}\IEEEauthorrefmark{2}, 
 Jiaan Wu\IEEEauthorrefmark{2}, 
 Zedong Liu\IEEEauthorrefmark{1}\IEEEauthorrefmark{2}, 
 Xinyang Ma\IEEEauthorrefmark{1}\IEEEauthorrefmark{2}, 
 Hairui Zhao\IEEEauthorrefmark{1}\IEEEauthorrefmark{2}, 
 Yida Gu\IEEEauthorrefmark{1}\IEEEauthorrefmark{2}, 
 Yuanhong Huang\IEEEauthorrefmark{1}\IEEEauthorrefmark{2},\\
 Xingchen Liu\IEEEauthorrefmark{1}\IEEEauthorrefmark{2},
 Wenjing Huang\IEEEauthorrefmark{1}\IEEEauthorrefmark{2}, 
 Zheng Wei\IEEEauthorrefmark{1}, 
 Jing Xing\IEEEauthorrefmark{1}, 
 Yili Ma\IEEEauthorrefmark{1}, 
 Qingyi Zhang\IEEEauthorrefmark{3}, 
 Baoyi An\IEEEauthorrefmark{3},\\
 Zhongzhe Hu\IEEEauthorrefmark{3}, 
 Shaoteng Liu\IEEEauthorrefmark{3},
 Xia Zhu\IEEEauthorrefmark{3}, 
 Jiaxun Lu\IEEEauthorrefmark{3}, 
 Guangming Tan\IEEEauthorrefmark{1}\IEEEauthorrefmark{2}, 
 and Dingwen Tao\IEEEauthorrefmark{1}\IEEEauthorrefmark{2}}

 	\IEEEauthorblockA{
 		\IEEEauthorrefmark{1}Institute of Computing Technology, Chinese Academy of Sciences, Beijing, China
 	}
 	\IEEEauthorblockA{
 		\IEEEauthorrefmark{2}University of Chinese Academy of Sciences, Beijing, China
 	}
 	\IEEEauthorblockA{
 		\IEEEauthorrefmark{3}Huawei Technologies Co., Ltd., Shenzhen, Guangdong, China
 	}
 \thanks{Dingwen Tao is the corresponding author (taodingwen@ict.ac.cn).}%
 }

\maketitle


\begin{abstract}
The rapid scaling of Large Language Models presents significant challenges for their deployment and inference, particularly on resource-constrained specialized AI hardware accelerators such as Huawei's Ascend NPUs, where weight data transfer has become a critical performance bottleneck. While lossless compression can preserve model accuracy and reduce data volume, existing lossless compression algorithms exhibit extremely low throughput when ported to the Ascend NPU architecture. In this paper, we propose ENEC, a novel lossless compression method specifically customized for AI model weights and optimized for Ascend Neural Processing Units. ENEC adopts a block-based fixed-length encoding scheme and incorporates a series of NPU-specific optimizations: bit-width quantization with hierarchical halving bit-packing, vectorized branch-free integer transformation, and dependency-decoupled intra-segment scan for efficient prefix-sum computation. Experimental results demonstrate that ENEC outperforms existing state-of-the-art NPU compressors in both compression ratio and throughput. Compared to leading GPU solutions, ENEC achieves a 3.43× higher throughput than DietGPU and a 1.12× better compression ratio than nvCOMP. By reducing weight transmission overhead, ENEC significantly improves end-to-end inference performance, achieving up to a 6.3× speedup. On Ascend NPUs, ENEC is the first open-source lossless compression algorithm for model weights that achieves performance comparable to state-of-the-art GPU compressors, offering an effective solution for deploying large-scale AI models.
\end{abstract}

\maketitle
\section{Introduction}\label{sec:introduction}
Large Language Models (LLMs) have demonstrated remarkable performance across diverse application domains~\cite{abdin2024phi, achiam2023gpt, dubey2024llama}, driving significant demand for efficient training and inference. In response, a range of specialized AI accelerators has been developed, including the Cerebras CS2~\cite{cerebras-doc}, SambaNova SN30~\cite{sambanova-doc}, and the Groq GroqChip~\cite{github-groqflow}. Among these, Huawei’s Ascend Neural Processing Units (NPUs) stand out due to their high computational throughput, strong energy efficiency, and favorable cost–performance characteristics compared to conventional GPUs. Notably, systems such as Huawei’s CloudMatrix384 platform, equipped with CloudMatrix-Infer~\cite{zuo2025serving}, demonstrate impressive scalability on large-scale AI workloads.

\begin{figure}[!t]
    \centering
    \includegraphics[width=1\linewidth]{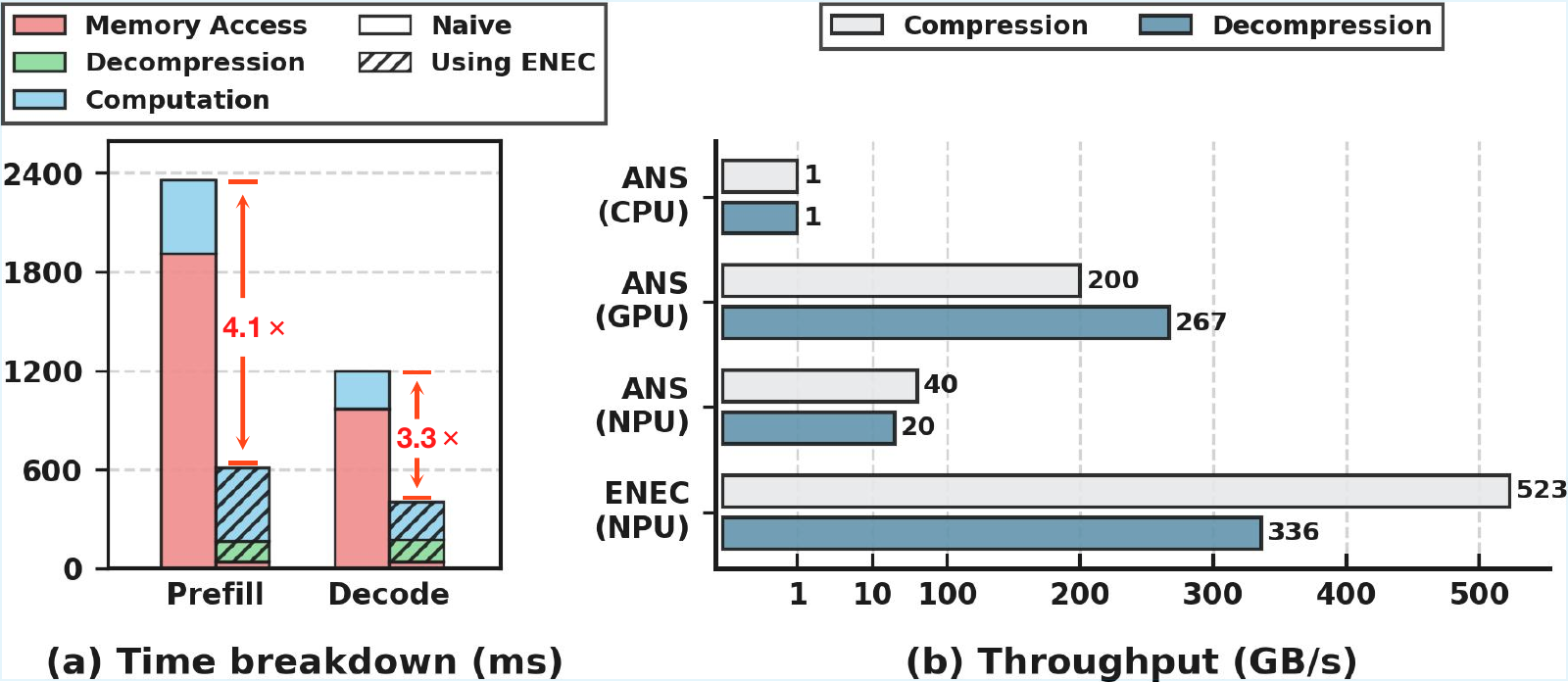}  
    \caption{\textcolor{black}{Time breakdown of Qwen3-32B Inference on NPU 910B2 (left) and throughput of ANS across multiple platforms (right).}}
    \label{fig:intro}
\end{figure}

Despite the hardware advances, the rapid growth in model size poses severe challenges for deployment and inference efficiency~\cite{yuan2024llm, zhou2024survey}, especially in resource-constrained environments. For instance, the Llama-3.1-405B model~\cite{dubey2024llama} contains 405 billion parameters and requires approximately 910 GB of memory for full inference in BFloat16 precision~\cite{deepseek-llm,bert,dubey2024llama,Jiang2023Mistral7,qwen3technicalreport}, exceeding the total HBM capacity of even high-end NPU servers equipped with 8 NPU 910B2s (each with 64 GB HBM). Consequently, deploying such massive models requires distributing model weights across CPUs and NPUs. The frequent CPU-NPU data transfers for accessing remote weights during computation have thus become a critical performance bottleneck~\cite{jia2024sdp4bit,jiang2024megascale,li2024multidimensional}. \textcolor{black}{
As Figure~\ref{fig:intro}(a) illustrates, memory access constitutes the dominant factor in inference latency for the Qwen3-32B model, accounting for 78\% to 85\% of the total execution time in both the prefill and decode stages~\cite{qwen3technicalreport}. This severe I/O bottleneck effectively marginalizes computation, drastically underutilizing the hardware. Consequently, this naive deployment strategy results in latencies that are 4.1× and 3.3× higher than our optimized ENEC approach for the prefill and decode stages, respectively. Such overhead not only intensifies the performance bottleneck but also escalates operational costs and hinders accessibility, ultimately limiting the practical deployment of state-of-the-art (SOTA) LLMs.}


To address the challenges associated with model scaling, numerous weight compression techniques have been proposed, including quantization~\cite{frantar2022gptq,lin2024awq,xiao2023smoothquant}, pruning~\cite{ashkboos2024slicegpt,frantar2023sparsegpt}, and low-rank decomposition~\cite{wang2024svd,yuan2023asvd}. These methods substantially reduce memory footprint and computational overhead, enabling faster inference under resource constraints. However, they are inherently lossy and often alter the model's output distribution, potentially compromising accuracy and reliability. In contrast, lossless compression techniques—such as entropy coding based on Asymmetric Numeral Systems (ANS)~\cite{duda2013asymmetric,pieprzyk2023compression,lin2023recoil,weissenberger2019massively,yamamoto2024asymptotic}—preserve full model fidelity by ensuring exact reconstruction of the original weights. Notable examples include general-purpose algorithms like Zstandard (Zstd)~\cite{collet2018zstandard} and nvCOMP~\cite{NVCOMPNV83:online}, as well as specialized encoders tailored for model weights, such as ZipNN~\cite{hershcovitch2025zipnn}, DFloat11~\cite{zhang202570}, and Huff-LLM~\cite{yubeaton2025huff}, all of which have demonstrated strong performance on conventional CPU or GPU platforms.


\textcolor{black}{Lossless compression for LLMs is particularly needed for Ascend NPUs because these accelerators—specifically engineered for AI models—are typically highly capable in terms of computational resources, such as matrix and tensor cores, while their HBM capacity and NPU–CPU bandwidth remain limited. Thus, effective lossless compression on NPUs can help bridge this gap.}
However, an open-source lossless compressor for model weight compression on NPUs is notably absent. To investigate this opportunity, we ported an ANS compressor—an algorithm known for combining the speed of Huffman coding~\cite{hidayat2023increasing,huffman2007method,iyer2025advancement,lu2023adt,shah2023lightweight,yamamoto2020huffman} with compression efficiency approaching that of arithmetic coding~\cite{langdon1984introduction}—from conventional CPU and GPU environments to the NPU. Unfortunately, our implementation achieved \textit{exceptionally poor throughput}, as shown in Figure~\ref{fig:intro}(b). This performance limitation is particularly critical in inference scenarios, where compressed weights must be decompressed in real time, making the decompression phase itself a severe system bottleneck. Moreover, the issue is not unique to ANS; similar throughput degradation was observed when porting other lossless algorithms such as LZ77~\cite{ziv2003universal}. \textcolor{black}{
We conclude that existing lossless compressors are fundamentally incompatible with Ascend architecture. The SIMD-based vectorized design lacks conditional branching, scatter/gather capabilities, and efficient variable-length memory handling—severely limiting parallelization and throughput of traditional compression algorithms, as further analyzed in Section~\ref{sec:architecture_analysis}.}

This paper proposes \underline{E}fficient \underline{N}PU-\underline{E}nhanced \underline{C}ompression (ENEC), a novel hardware-software co-designed algorithm for model weights. Designed to harness the full capabilities of Ascend NPUs, ENEC uses a block-based, fixed-length approach with NPU-specific optimizations like bit-width quantization with hierarchical halving bit-packing that boosts compression speed without sacrificing the compression ratio; a table-based mapping strategy combined with a branch-free integer transform, which replaces memory accesses with arithmetic computations to optimize the statistical characteristics of model weights; and an intra-segment dependency decoupled scan for efficient prefix sum. ENEC attains a compression throughput ranging from 263 to 523 GB/s and a decompression throughput of 188 to 336 GB/s, outperforming state-of-the-art NPU compressors by up to 2.47× and 2.11× while maintaining high compression ratio. This breakthrough substantially reduces data transmission overhead, enabling up to a 6.3× speedup in end-to-end inference.

Our contributions are summarized as follows:
\begin{itemize}[topsep=2pt,leftmargin=1em]
\item We obtain several key observations through in-depth analyses of various AI model weights, which informed our basic design and optimizations.
\item We design an efficient lossless encoding method for Ascend NPUs that overcomes architectural constraints, based on our comprehensive analysis of lossless compression incompatibility and an extension of the LC framework.
\item We propose a set of optimizations to achieve both high compression ratios and high throughput on Ascend NPUs, including bit-width quantization with hierarchical halving bit-packing, vectorized branch-free integer transformation, and intra-segment dependency–decoupled scan.
\item 
Evaluated on Ascend 910B2 with 10 real-world AI models, ENEC outperforms SOTA NPU compressors in compression ratio and throughput. Compared to GPU compressors, ENEC achieves 3.43× higher throughput than DietGPU and a 1.12× better ratio than nvCOMP, yielding up to a 6.3× end-to-end inference speedup.
\end{itemize}

\textit{To the best of our knowledge}, \textcolor{black}{ENEC\footnote{\textcolor{black}{Our code is available at \url{https://github.com/hpdps-group/ENEC}.}} is the first open-source compressor for AI model on Ascend NPUs} that achieves performance comparable to SOTA GPU compressors while maintaining the model weight compression ratio.

\begin{figure}[t]
    \centering
    \includegraphics[width=1.0\linewidth]{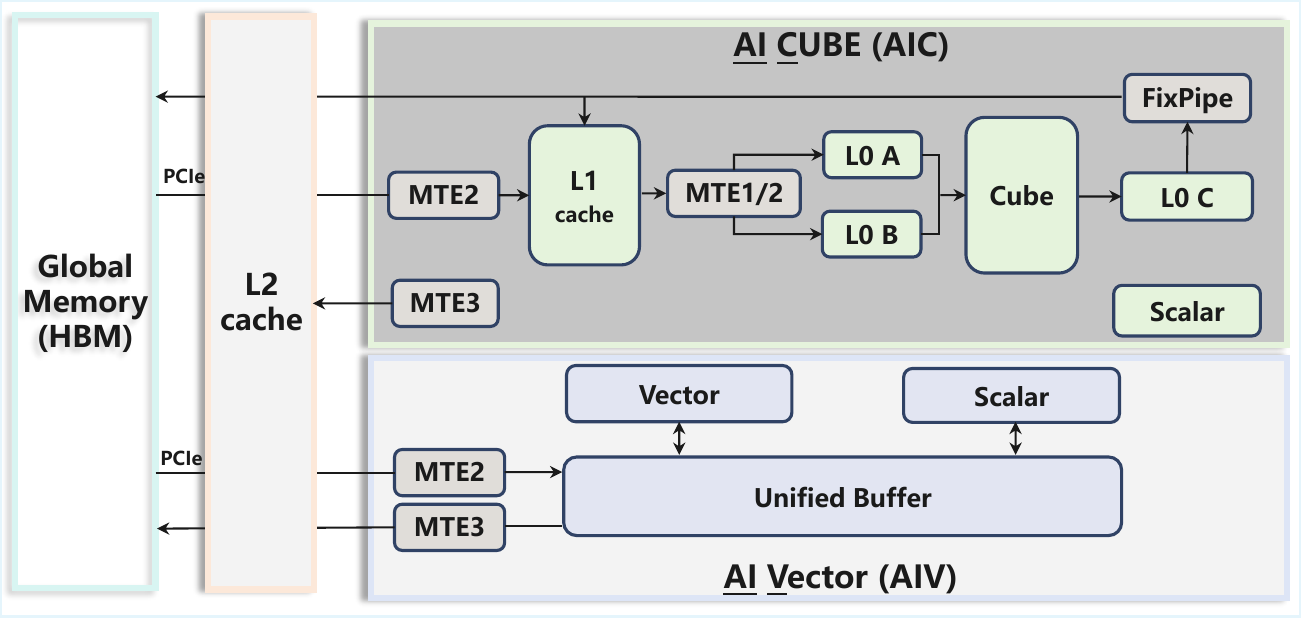}  
    \caption{DaVinci architecture (decoupled AIC and AIV).}
    \label{fig:npu_arch}
\end{figure}

\section{Background}
\label{sec:background}

\subsection{Ascend NPUs}
\label{ssec:npu_architecture}

The Ascend NPUs are specialized accelerator chips built on the DaVinci architecture~\cite{liao2019davinci} for deep learning workloads.

\textbf{Compute Architecture and Core Specialization.} 
The Ascend NPUs consist primarily of memory units and computing units. The main computing components are AI Cores. One AI core typically integrates one AI Cube (AIC) unit and two AI Vector (AIV) units. For instance, the Ascend 910B2 NPU incorporates 24 such AI Cores, providing a total of 24 AIC units and 48 AIV units. The AIC is optimized for high-performance matrix operations, offering exceptional throughput for dense linear algebra, albeit with limited operational flexibility. In contrast, the AIV supports large-scale vectorized operations, including data gathering, reduction, and other collective operations.

\textbf{Memory Hierarchy and Data Path.} On the memory side, the Ascend NPUs feature a high-bandwidth memory (HBM) and an L2 cache shared across all AI Cores. For AIV operations, input data must first be loaded into a local buffer known as the Unified Buffer (UB), typically on the order of kilobytes (e.g., 192KB). The AI Cube unit employs multiple levels of on-chip buffers such as L1, L0A, L0B, L0C, BT, and FP buffers, as illustrated in Figure~\ref{fig:npu_arch}. Data movement between different memory hierarchies is managed by Memory Transfer Engines (MTEs), which not only handle data transfers but can also perform in-flight data format and type conversions.

\textbf{Programming Model and Execution Pipeline.} Huawei introduced AscendC, a C++ programming model within Compute Architecture for Neural Networks (CANN)~\cite{cann-doc}, to enable high-performance operator development on Ascend NPUs. AscendC employs a multi-pipeline abstraction to enable fine-grained hardware control, allowing developers to fully exploit hardware's potential. It simplifies programming through key abstractions like tensors and queues. A tensor encapsulates data that will be operated on by the AIC/AIV units, while queues handle synchronization: after an operation completes, the result is enqueued (EnQue), and dependent operations dequeue (DeQue) it. More details can be found in the user documentation~\cite{cann-doc}.


\subsection{Lossless Floating-Point Compressors}\label{sec:grad_comp}
\textbf{General-purpose Lossless Floating-point Compressors.} 
Lossless compressors for floating-point data enable compression without information loss, making it valuable for applications like image~\cite{du2009novel}, time series data~\cite{li2023elf,li2025adaptive}, and scientific data~\cite{azami2025efficient,chen2023fcbench,knorr2021ndzip}.
The core techniques of lossless compression primarily fall into two categories: one leverages the identification of spatial redundancy patterns, while the other exploits the varying frequencies of different symbols. The first category is characterized by LZ-based compression methods (e.g., LZW~\cite{welch1984technique}, LZ78~\cite{ziv2003compression}, LZ77~\cite{ziv2003universal}) and related optimizations~\cite{zhang2023gpulz}, which operate by sequentially scanning the data stream and replacing repeated substrings with shorter references. The second category is represented by entropy coding techniques such as Huffman coding~\cite{hidayat2023increasing,huffman2007method,iyer2025advancement,lu2023adt,shah2023lightweight,yamamoto2020huffman} and arithmetic coding~\cite{langdon1984introduction}, which assign shorter bit representations to more frequently occurring symbols, thereby reducing the overall bit length. 

\textbf{Dedicated Lossless Floating-point Compressors for AI Model.} Despite the high compression efficiency and throughput of general-purpose compressors like Zstd~\cite{collet2018zstandard} on typical data types, their performance often degrades significantly on model weight data. Recently, several methods (e.g., ZipNN~\cite{hershcovitch2025zipnn} and its extension~\cite{heilper2025lossless}, DFloat11~\cite{zhang202570}, Huff-LLM~\cite{yubeaton2025huff}) have identified that the high randomness of floating-point mantissas interferes with the compressibility of exponents, leading to suboptimal results. Moreover, DietGPU~\cite{github-dietgpu}, the ANS-based~\cite{duda2013asymmetric,pieprzyk2023compression,lin2023recoil,weissenberger2019massively,yamamoto2024asymptotic} compressor on GPU, has introduced a dedicated float codec specifically designed for model floating-point data. To address this, they propose separating the exponent and mantissa components before compression. While this approach improves compression ratios, the exponent compression still relies on variable-length coding, which is poorly suited for efficient implementation on Ascend NPUs due to irregular memory access and control flow, thereby limiting their practical deployment efficiency on Ascend NPUs. Recently, Huawei has developed a compression algorithm named HANS~\cite{hans-doc}; however, its compression ratio and throughput performance still remain limited and it is also closed-source.

\subsection{LC Framework to Search Optimal Compression Algorithm}
\label{ssec:lc_framework}

To achieve efficient compression on Ascend NPUs, lightweight operations are essential. For this purpose, we have enhanced the emerging and significant open-source data compression tool—the LC framework~\cite{github-lcframework} and utilized it to search for lightweight component combinations that deliver either the optimal compression ratio or the highest throughput. This framework provides a wide range of common compression components and preprocessing methods, among which the Reducer is the sole component used for shortening data sequences, employing techniques such as HCLOG, RLE, RRE, and RZE to leverage various types of data redundancy.

Among these techniques, HCLOG employs a grouped bit-packing strategy: it divides a 16 KB data block into a fixed number of 32 sub-chunks, calculates the minimum leading zero count in each sub-chunk as metadata, and stores only the valid bits. However, a single outlier can force the entire sub-chunk to adopt a higher bitwidth, thereby reducing the compression ratio. To address this issue, we have extended the LC framework with a series of HCLOG compressors that support varying numbers of sub-chunks.

\section{Data Analysis of AI Model Weights}
\label{sec:analysis}


\textcolor{black}{
Model weights are typically stored in one of three floating-point formats: FP32, FP16, or BF16. Recently, BF16 has gained popularity for large models due to its wider exponent range, which offers training stability~\cite{kalamkar2019study, xi2025coatcompressingoptimizerstates, lee2025fp8againquantifyingreduced}. Although lower-precision integer formats such as INT8~\cite{dettmers2022llmint88bitmatrixmultiplication} and FP8~\cite{micikevicius2022fp8formatsdeeplearning} have been explored for model compression, they are not the primary focus of this study for several reasons. First, INT8 and FP8 offer lower precision than BF16, risking accuracy. Second, INT8 and FP8 formats typically require quantization, which constitutes a separate lossy compression step and offers limited additional compressibility beyond the quantization itself. Third, while the Ascend NPU's Cube units provide hardware support for INT8, the AIV API—the core framework of our implementation—lacks the necessary programming interfaces and instruction sets for the bit-manipulation and transformation tasks required by our method. Moreover, FP8 is entirely unsupported by the hardware, further precluding its use. Since models in other floating-point formats exhibit similar statistical properties, our analyses focus on BF16 as a representative case.
}

\begin{table}[htbp]
    \caption{LC search results. \textcolor{black}{``Others'' represents the set of all non-HCLOG algorithmic variants within the LC framework.}}
    \resizebox{\linewidth}{!}{    
        \begin{tabular}{l|c|c|c}
            \toprule[0.5mm] 
            \multirow{2}{*}{\textbf{LC-Alg}} & \multicolumn{3}{c}{\textbf{Models}} \\
            \cmidrule(lr){2-4}
            & deepseek-llm-7b-base  \cite{deepseek-llm} & Qwen3-8b  \cite{qwen3technicalreport} & Llama-3.1-8B-Instruct  \cite{dubey2024llama} \\
            \midrule[0.4mm]
            \textbf{HCLOG} & \textbf{98.23\%} & \textbf{98.30\%} & \textbf{99.36\%} \\
            Others & 1.77\% & 1.70\% & 0.64\% \\
            \bottomrule[0.5mm]
        \end{tabular}
    }
    \label{tab:lc_results}
\end{table}

\textbf{Observation 1: The exponent is more compressible than the sign and mantissa.} \label{obs:1}
BF16 format uses 1 sign bit, 8 exponent bits, and 7 mantissa bits. Analysis of DeepSeek weights shows sign and mantissa bits have uniform distribution, while exponents are highly non-uniform. Entropy calculations confirm: sign and mantissa have high entropy (7.97 bits), making compression difficult; exponents show low 2.58-bit entropy with strong compression potential. This insight aligns with findings from recent studies such as ZipNN~\cite{hershcovitch2025zipnn} and DFloat11~\cite{zhang202570}.

\textbf{Observation 2: In \textcolor{black}{the} LC framework, the HCLOG variant achieves the highest compression ratio in most cases.} \label{obs:2}\textcolor{black}{We use model weight tensors as input. Specifically, we partition the large-scale model parameters into multiple smaller files to enable a more granular analysis. For each file, we independently invoke the improved LC framework~\cite{github-lcframework} to search for its optimal configuration, thereby identifying a lightweight compressor that maximizes compression ratios.} As shown in Table \ref{tab:lc_results}, results across various model weight datasets consistently show the HCLOG variant achieves the highest ratio in most cases.

\textbf{Observation 3: Exponent values are highly constrained, concentrated within a narrow continuous range.} \label{obs:3}
Our further analysis reveals that the actual range of exponent values is highly constrained, with many potential values completely absent from the weights. Comprehensive examination demonstrates that these exponent values are concentrated within a narrow continuous range.


\textbf{Observation 4: The bit widths of the data grouped by the HCLOG grouping method after mapping by frequency sorting are obviously consistent.} \label{obs:4}By applying frequency-based mapping and using the grouping method from the HCLOG approach mentioned in Section \ref{ssec:lc_framework}, we observe that most grouped data has a bit width $\leq$ m, while a small portion has a bit width $>$ m.

\begin{figure}[h]
    \centering
    \includegraphics[width=0.8\linewidth]{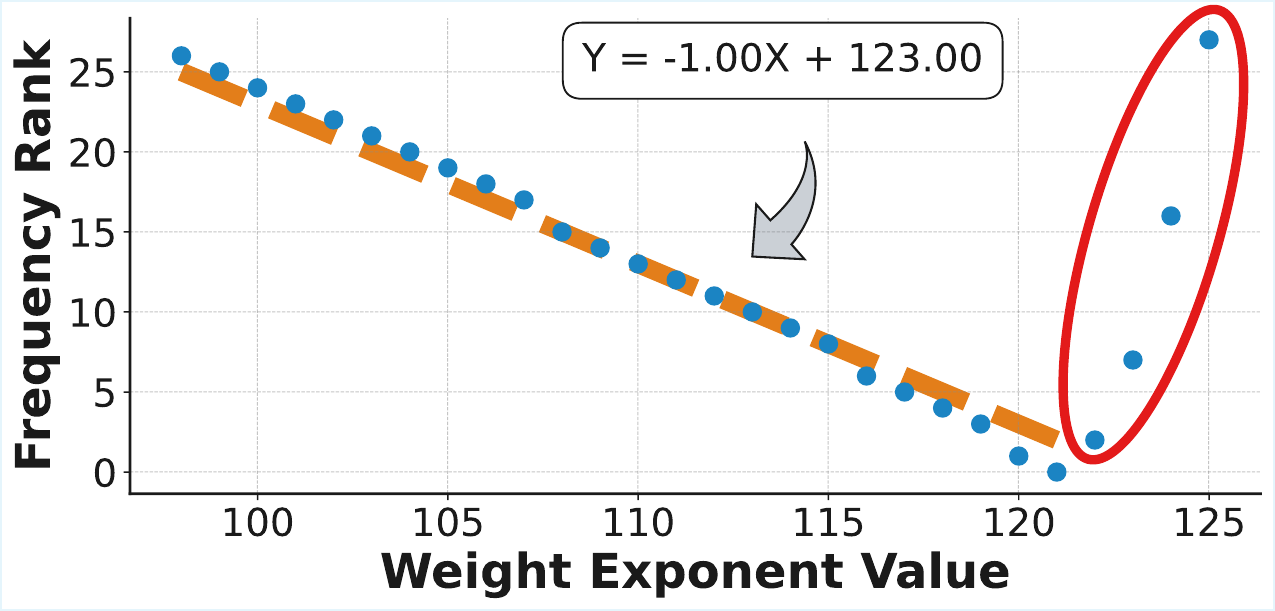}  
    \caption{Linear relationship between exponent values and frequency rankings in model weights. The red circle indicates outliers.}
    \label{fig:linear_fitting_plot}
\end{figure}

\textbf{Observation 5: A significant linear relationship exists between the weight exponent values and their frequency rankings.} \label{obs:5} We conduct an in-depth analysis of the frequency distribution of weight exponent values and observed a distinct negative linear relationship between the exponent values and their frequency ranking. As shown in Figure~\ref{fig:linear_fitting_plot}, this relationship can be effectively fitted by a linear function such as $Y = -1.00X + 123.00$, where the X-axis represents the exponent value and the Y-axis represents the corresponding frequency ranking (a lower ranking indicates higher frequency). This highly regular and predictable distribution characteristic serves as a critical foundation for efficient compression of model weights. It is worth noting that a small number of high exponent value points marked by the red elliptical region in the figure deviate from the main linear trend, representing rare outliers. 

\begin{figure}[h]
    \centering
    \includegraphics[width=0.9\linewidth]{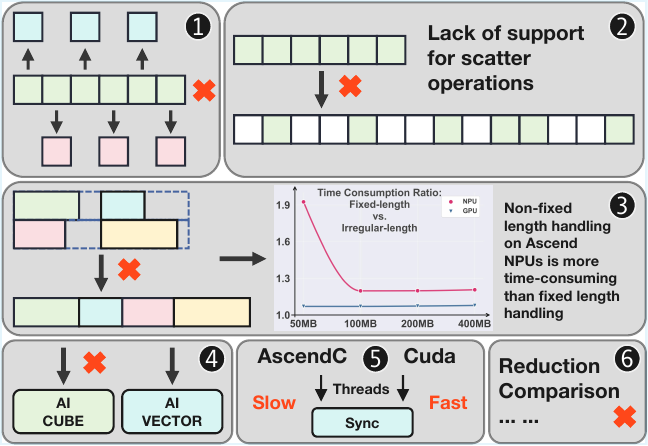}  
    \caption{\textcolor{black}{Execution challenges and constraints on Ascend NPUs.}}
    \label{fig:Analysis}
    \vspace{-2mm}
\end{figure}

\section{Architecture Analysis and Basic Design}
\label{sec:basic}

\subsection{Architecture Analysis of Lossless Compression Incompatibility on Ascend NPUs}
\label{sec:architecture_analysis}

As Figure~\ref{fig:Analysis} shows, the fundamental incompatibility between Ascend NPUs and lossless compression algorithms stems from architectural mismatches at multiple levels:
\begin{enumerate}[label=\protect\mycircle{\arabic*}, topsep=2pt, leftmargin=2em, labelsep=0.5em]
\item \textcolor{black}{The current Ascend architecture, based on SIMD, does not support flexible element-wise operations such as conditional if branching. Furthermore, it has limited instructions for integer arithmetic, making it particularly unsuitable for entropy coding.}
\item \textcolor{black}{The lack of instructions for converting between contiguous and non-contiguous memory layouts, such as \texttt{scatter}, severely degrades decoding performance.}
\item Handling and transferring variable-length memory operations is inefficient and time-consuming, severely impacting algorithms like Huffman coding.
\item Compression computations are primarily vector-based element-wise operations and cannot directly utilize the Ascend's Cube unit designed for matrix operations. Therefore, we construct a collaborative inference pipeline where compression is handled by the AIV and model operations are executed by the AIC.
\item \textcolor{black}{AscendC lacks fast inter-thread synchronization comparable to CUDA, as Ascend NPUs treat each AI Core as a single, heavyweight thread. The architecture relies on a pipelined dataflow model, optimizing task-queue-driven overlaps between data movement and computation within a thread, rather than managing fine-grained concurrency across massive threads.}
\item It is challenging to implement lossless compression algorithms that require extensive non-arithmetic operations (such as comparisons and reductions), including Run-Length Encoding (RLE), LZ4, LZ77, and other similar dictionary-based methods.
\end{enumerate}

\textcolor{black}{These architectural mismatches explain the fundamental incompatibility between existing lossless compression algorithms and Ascend NPUs. The lack of conditional branching, scatter/gather instructions, and efficient variable-length memory operations renders traditional entropy coding and dictionary-based methods impractical on this platform. Moreover, the absence of lightweight inter-thread synchronization and reliance on a pipelined dataflow model within each AI core restrict parallelization strategies common in GPU-based compressors. These insights motivate our co-design approach: rather than force-fitting existing algorithms, we align compression primitives with Ascend's vectorized, branch-free execution model and memory hierarchy constraints. Building on this understanding, we next present ENEC—a lightweight compression algorithm crafted to navigate these architectural limitations while preserving lossless fidelity.
}

\subsection{Basic Design of ENEC}
\label{sec:basic_design}
Building upon the architectural constraints identified in Section \ref{sec:architecture_analysis}, we now present the basic design of ENEC—a lightweight compression algorithm, as shown in the pseudocode (Algorithm \ref{alg:basic_design}). While the architectural analysis reveals fundamental incompatibilities, it also illuminates the path forward: by aligning our algorithm with the Ascend's inherent strengths and working around its constraints, we can achieve efficient lossless compression.

\textcolor{black}{
Guided by Observations 1 and 2 from Section \ref{sec:analysis}, our fundamental compression algorithm proceeds as follows. During the compression phase, each thread processes a data block of 8,192 elements at a time. The exponents are separated from the floating-point data (while signs and mantissas are stored directly), and frequency statistics are collected (Line 2). The sorted frequency indices serve as mapping values to achieve more compact encoding (Line 3). Within this 8,192-element block, data is further divided into smaller groups of length L, and the bit width required for the maximum value in each group is calculated (Lines 5-8). Each thread utilizes a dedicated 32KB buffer (with 8,192 32-bit lanes corresponding to 8,192 elements), where each element is written into its respective lane. As the number of processed blocks increases, the lower 16 bits of all lanes that have accumulated full 16 bits are gathered to output and a bit mask is generated, followed by a vectorized right-shift operation to prepare for the next round (Lines 10-16). This iterative process continues until all data blocks are fully processed. Finally, the compressed content from each thread's buffer is saved into the compressed file for decoding purposes.
}

\begin{algorithm}[t]
\scriptsize\sffamily
\normalfont
\DontPrintSemicolon 
\LinesNumbered 
\SetAlgoLined
\SetKwInOut{Input}{Input}
\SetKwInOut{Output}{Output}
\SetCustomAlgoRuledWidth{0.45pt}

\newcommand{\rbnode}[1]{\tcp*[r]{\makebox[3.0cm][l]{\color{red}{\textbf{#1}}}}}
\newcommand{\mycomment}[1]{\tcp{\textrm{#1}}}

\Input{$W$: Raw Weights (BF16/FP16/FP32), $T_{freq}$: Frequency Table}
\Output{$Stream$: Compressed Binary Output}
\BlankLine

\textbf{Compression (Processing 8,192 elements per block):} \\
\mycomment{Component Separation}
$\{E, S, M\} \gets$ Split($W$) \\

$E' \gets T_{freq}[E]$ \rbnode{[B1] Gather Lookup}

{\tiny (\textbf{\textcolor{black}{Limit}: High latency in non-contiguous memory access, ~35\% overhead})}

\mycomment{Group-based $bit$-$width$ calculation}
Divide $E'$ into groups of length $L$ \\
\For{each group $G$}{
    $bit$-$width$ $\gets \lceil \log_2(\max(G)) \rceil$ \rbnode{[B2] Reduction Max}
}

{\tiny (\textbf{\textcolor{black}{Limit}: Serial dependency in vector units, ~40\% overhead})}

\mycomment{Bit-packing} 
\While{bits remaining in $E'$}{
    \mycomment{Buffer Accumulation}
    Fill 8,192 lanes in 32KB $buffer$ \\
    \If{lane status $\ge$ 16 bits}{
        Output $lower$ $16$ $bits$ and $bit$ $mask$ to $Stream$ \\
        $lanes \gets lanes \texttt{ >> } 16$ 
    }
}
\BlankLine
\textbf{Decompression:} \\
\mycomment{Offset calculation and bit-unpacking} 
$Mask \gets$ Convert $bit$ $mask$ to $\{0, 1\}$ integers \\
$Offsets \gets$ PrefixSum($Mask$) \rbnode{[B3] Prefix Sum}

{\tiny (\textbf{\textcolor{black}{Limit}: Cross-lane data dependency within 32B segments, ~30\% overhead})}

$Lower$ $16$ $bits$ $\gets$ Get from $Stream$ using $Offsets$

$E' \gets$ $buffer$ $+$ $Lower$ $16$ $bits$

$E \gets T_{freq}^{-1}[E']$ \rbnode{[B4] Gather Lookup}
{\tiny (\textbf{\textcolor{black}{Limit}: Inefficient vector unit utilization in gather, ~45\% overhead})}

$W \gets$ Combine($E, S, M$) \\
\Return $W$

\caption{\footnotesize \textbf{\textcolor{black}{Basic Design of ENEC}} \protect\\ 
\textcolor{black}{The right-side markers [B1-B4] represent the primary architectural bottlenecks on Ascend NPUs addressed in Section \ref{sec:design}}}
\label{alg:basic_design}
\end{algorithm}

\begin{figure*} [htbp]
  \includegraphics[width=\linewidth]{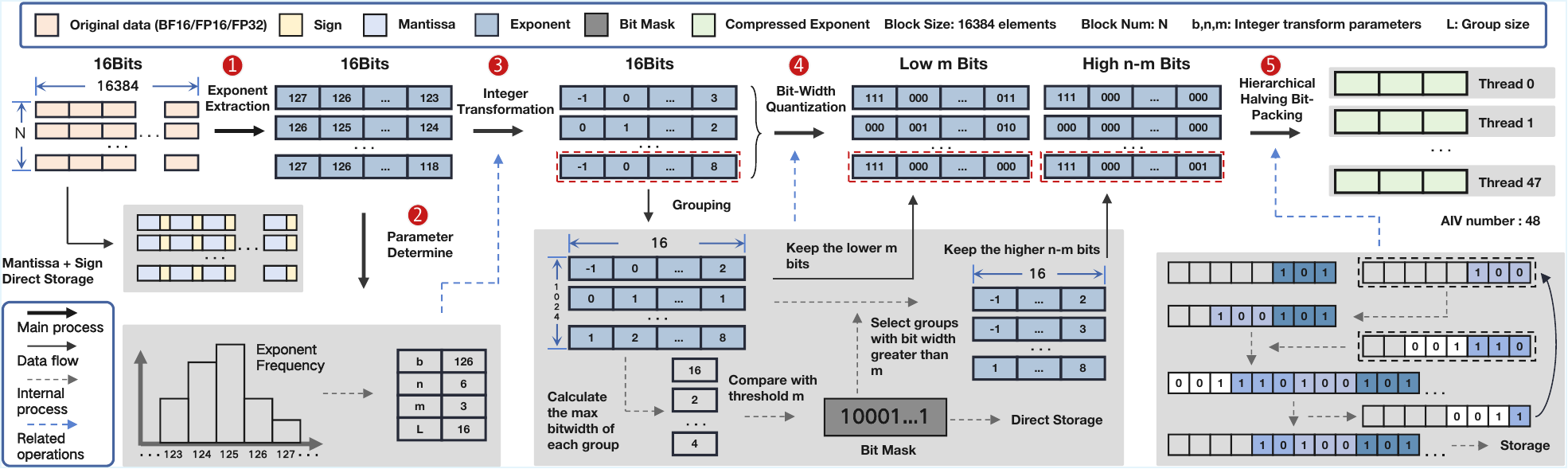}
  \caption{\textcolor{black}{Overview of optimized ENEC compression design on Ascend NPUs.}}
  \label{fig:overview}
  \vspace{-4mm}
\end{figure*}

\begin{figure}[t]
    \centering
    \includegraphics[width=1.0\linewidth]{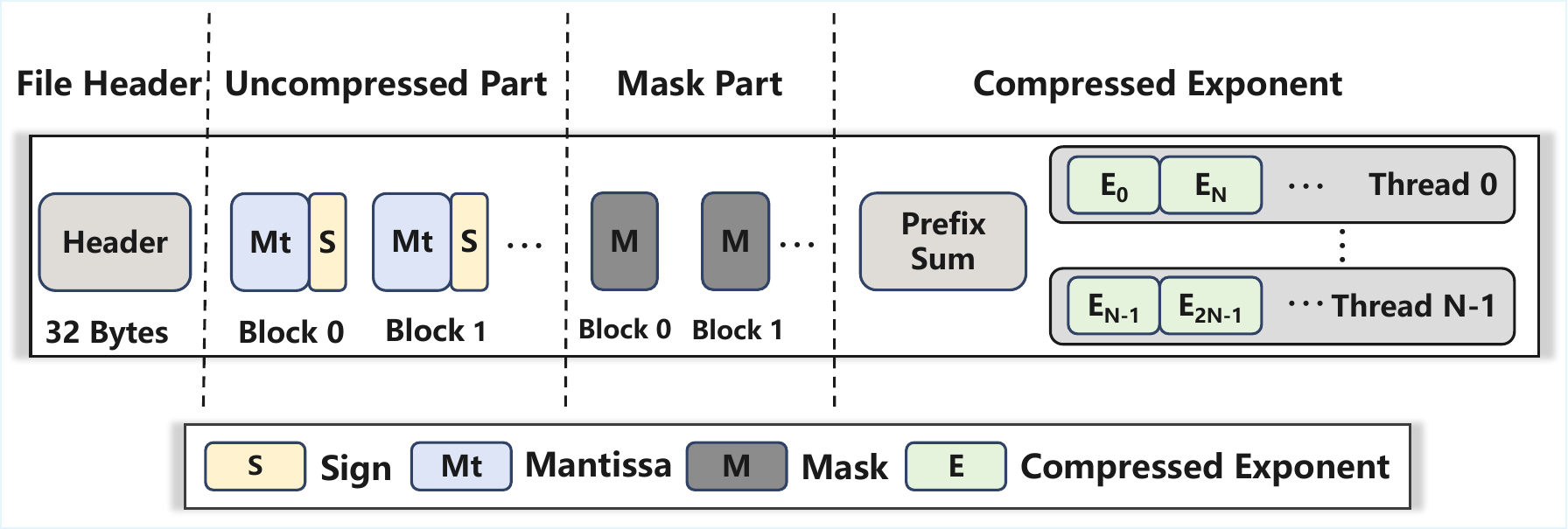} 
    \caption{Compressed stream layout. The bit mask is used to distinguish anomalous groups within each data block. Prefix sums provide direct starting positions for each thread’s compressed data.}
    \label{fig:stream_layout}
\end{figure}

\textcolor{black}{
During the decompression phase, bit-unpacking operations are first performed to reconstruct the grouped data. The inverse bit-packing employs a reverse gather operation, where the necessary offsets for the reverse gather are precisely computed using a prefix sum, which is derived by first converting the bit mask into a sequence of 0 and 1 integers (Lines 18-22). Subsequently, by executing a lookup in the frequency mapping table, the values are converted back to their original exponent values (Line 23). By combining these restored exponents with the directly stored sign and mantissa bits, the original floating-point data can be perfectly reconstructed (Line 25).
}

\textbf{Key Bottlenecks of Basic Design.} However, the basic design does not deliver optimal performance. Performance analysis of the compression kernel reveals that gather table lookups account for 35\% of the time (Line 3), while reduction max operations consume 40\% (Line 7). Similarly, analysis of the decompression kernel shows that prefix sum operations take 30\% of the time (Line 19), with gather table lookups occupying 45\% (Line 23). \textcolor{black}{Collectively, these limitations result in low throughput, preventing the system from achieving high-performance efficiency.} Therefore, in Section \ref{sec:design}, we comprehensively optimize the compression and decompression processes of the basic design based on Observations 3, 4 and 5 from Section \ref{sec:analysis}.

\section{Optimizations of {\thiswork}}
\label{sec:design}
\textcolor{black}{
Drawing on the observations from Section \ref{sec:analysis} and hardware characteristics from Section \ref{sec:basic}, this section details the optimizations on the basic ENEC from Section \ref{sec:basic_design} in their order of implementation, directly corresponding to the ablation studies presented in Section \ref{sec:evaluation}. Specifically, we first present the optimized ENEC algorithm in Section \ref{sec:overview}. Subsequently, from Section \ref{sec:bit_packing} to \ref{ssec:design_hybrid}, we delve into the specific optimization methods and fine-grained tuning mechanisms employed to maximize performance.
}

\subsection{\textcolor{black}{Overview of optimized ENEC}}
\label{sec:overview}
Our optimized ENEC approach, as illustrated in Figure \ref{fig:overview}, processes model weights by dividing them into fixed-length data blocks that are processed in parallel by multiple threads (with the number of threads determined by the number of AIV cores) in a cyclic manner.

During compression, the process begins by extracting exponent(\circlednum{1}) and calculating the exponent frequency to determine optimal parameters(\circlednum{2}). After applying the branch-free integer transformation(\circlednum{3}) in Section \ref{sec:branch_free}, the data is grouped and the bit-width of each group's maximum value is determined. This bit-width is compared against a threshold to generate and store a compact bit mask and then it distinguishes elements based on this bit mask(\circlednum{4}). Finally using hierarchical halving bit-packing(\circlednum{5}) in Section \ref{sec:bit_packing} to generate compressed file (Figure \ref{fig:stream_layout}).

During decompression, the bit mask is first read and converted to integer values, followed by computing prefix sums using the intra-segment dependency decoupled scan (Section~\ref{sec:idd_scan}) to obtain offset for the reverse gather operation on the compressed data stream. Subsequently, an inverse vectorized, branch-free integer transformation is performed to restore the exponent data. Finally, the original data is reconstructed by reading the stored sign and mantissa bits.

\subsection{Bit-Width Quantization with Hierarchical Halving Bit-Packing}
\label{sec:bit_packing}
In the basic design, the variable bit-width packing requires inefficient reduction max and multiplication/division operations on Ascend NPUs. To address this, based on the observation 3 and 4 in Section \ref{sec:analysis}, we adopt the lossless bit-width quantization with hierarchical halving bit-packing techniques. The data block is organized using a group-interleaved scheme, where each group comprises $L$ elements. The number of bits required to pack a group is determined by calculating the bit width of the largest element within that group. We propose a two-level bit-width quantization strategy: if the required bit width is less than or equal to a threshold $m$, all elements within the group are stored uniformly using $m$ bits; otherwise, all elements are stored using $n$ bits, where $n$ represents the minimum number of bits required to represent different occurring exponents. This approach enables substituting the computationally expensive reduction max operation with efficient bitwise OR operation.

\setlength{\textfloatsep}{0pt}
\begin{algorithm}[t]
\scriptsize\sffamily
\normalfont
\DontPrintSemicolon
\SetAlgoLined
\LinesNumbered
\setcounter{AlgoLine}{0}
\SetKwInOut{Input}{Input}
\SetKwInOut{Output}{Output}

\newcommand{\mycomment}[1]{\tcp{\textrm{#1}}}

\Input{$data[0..N-1]$: Array of $N$ 16-bit data values ($N=2^k$); \\
$a$: Bit-width of each input value ($0 < a \leq 8$)}
\Output{$byte\_stream$: Compressed byte-aligned output stream}
\BlankLine

\mycomment{Initialization of packing parameters}
$total\_length \gets 0$ \\
$width \gets a$ \\
$length \gets N$ \\
$normalized\_bytes \gets$ empty vector \\

\While{$width > 0$}{
    \mycomment{Hierarchical halving: doubling width by merging pairs}
    \While{$length > 1$ \textbf{and} $width < 8$}
    {
        $length \gets length / 2$ \\
        \For{$i \gets 0$ \textbf{to} $length - 1$}{
            $data[i] \gets data[i] \mid (data[i + length] \texttt{ << } width)$ \\
        }
        $width \gets width \times 2$ \\
    }
    
    \mycomment{Extract aligned bytes from the packed 16-bit lanes}
    \For{$j \gets 0$ \textbf{to} $length - 1$}{
        \mycomment{Mask out the lower 8 bits and update residual data}
        $temp\_bytes[j] \gets (data[j] \texttt{ << } 8) \texttt{ >> } 8$ \\
        $data[j] \gets data[j] \texttt{ >> } 8$ \\
    }
    Append $temp\_bytes$ to $normalized\_bytes$ \\
    
    $width \gets width - 8$ \\
    $total\_length \gets total\_length + length$ \\
}

\mycomment{Padding to ensure even length for 16-bit aligned output}
\If{$total\_length \% 2 \neq 0$}{
    Append $0$ to $normalized\_bytes$ \\
    $total\_length \gets total\_length + 1$ \\
}

\mycomment{Final stream construction via vectorized concatenation}
\For{$i \gets 0$ \textbf{to} $total\_length / 2 - 1$}{
    $output\_data \gets normalized\_bytes[i] \mid (normalized\_bytes[i + total\_length / 2] \texttt{ << } 8)$ \\
    Append $output\_data$ to $byte\_stream$ \\
}

\Return $byte\_stream$ \\

\caption{\footnotesize \textbf{\textcolor{black}{Hierarchical Halving Bit-Packing with Byte Normalization (Vectorized)}}}
\label{alg:bit_packing_vectorized}
\end{algorithm}

To achieve both extreme compression and high performance for tensor blocks under strict hardware alignment constraints, we introduce an iterative bit-packing algorithm. The core of this method, as shown in the pseudocode (Algorithm \ref{alg:bit_packing_vectorized}), lies in lane folding and byte normalization. The algorithm begins with a block of $N$ elements (where $N$ is a power of two), each of 16-bit width but retaining only the least significant $a$ bits. It iterates until the effective bit width of each element is reduced to 8 bits. In each iteration, the block is treated as a sequence of logical lanes. The folding step logically left-shifts each element in the lower half by $a$ bits and performs an element-wise Or operation with the corresponding element in the upper half (Lines 9-11). This merges two $a$-bit payloads into a single physical storage location, resulting in a new block of $N/2$ elements, each with an effective width of $2a$ bits (Line 12).

When the folding operation causes the effective bit width to exceed the 8-bit byte boundary, \textit{byte normalization} is triggered. As the folding is iterative, this effective bit width grows with each iteration, potentially surpassing 8 bits multiple times. The data is then split into (i) the lower 8 bits, forming a storable byte, and (ii) the remaining bits, which are treated as overflow. All overflow segments are collected into a new sub-block and processed recursively by the same algorithm (Lines 14-21). Then, by appending zero bytes to the end of the byte stream, the total length is aligned upward to the nearest multiple of 2 (Lines 23-26). The normalized bytes are consolidated in a final folding pass to form the output stream (Lines 28-32). The bit-unpacking process is precisely the inverse of the bit-packing process.

Consequently, \textcolor{black}{as illustrated in Figure \ref{fig:overview},} the least significant $m$ bits are truncated and compressed via hierarchical halving bit-packing. For groups needing more than $m$ bits, the higher $(n - m)$ bits are collected into a 32 KB buffer. Once full, these bits are compressed by hierarchical halving bit-packing, and the buffer is then cleared. During decompression, an inverse \texttt{Gather} operation redistributes the decompressed higher $(n - m)$ bits to their original positions. The original values are reconstructed by performing vectorized Or operation with the decompressed lower $m$ bits.

\subsection{Vectorized Branch-Free Integer Transformation} 
\label{sec:branch_free}

\begin{figure}[t]
    \centering
    \includegraphics[width=\linewidth]{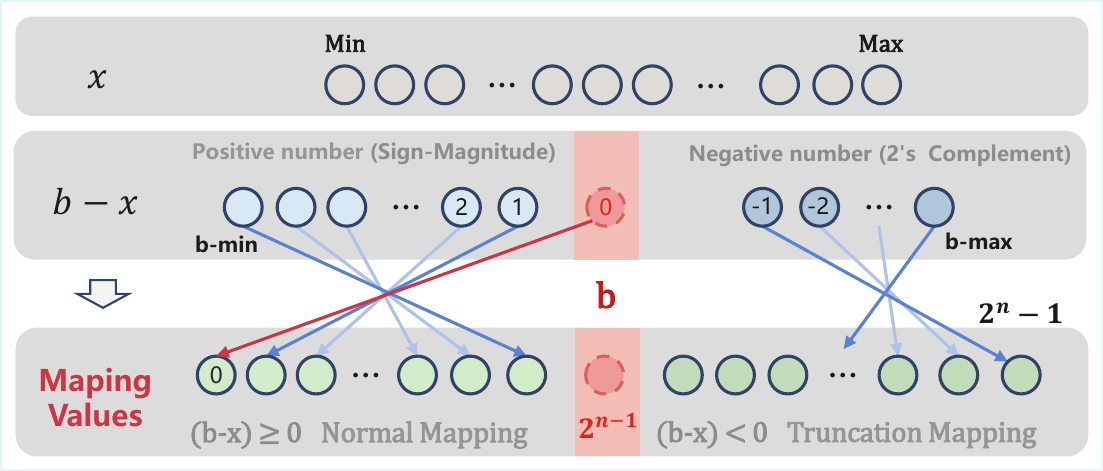}  
    \caption{Vectorized branch-free integer transformation.}
    \label{fig:vector_branchfree}
\end{figure}

In the basic design, the exponent mapping relies on the slow gather API, based on the observation 5 in Section \ref{sec:analysis}, we adopt the linear mapping function \(f(x) = b - x\) as the core transformation scheme to optimize this process. As shown in Figure \ref{fig:vector_branchfree}, \(b\) is a linear mapping parameter determined by the data distribution characteristics. When the input value is greater than the parameter \(b\), the mapping result is negative. This approach cleverly utilizes the properties of two's complement representation to handle negative values. Due to the use of bit-width quantization and the need to ensure correct decoding, an additional bit is required in the higher bit position to determine whether the value is negative: for positive values, the \((n+1)\)-th bit must be 0, and the mapping range is \([0, 2^n - 1]\); for negative values, the \((n+1)\)-th bit must be 1, and the mapping range is \([2^n, 2^{n+1} - 1]\). This approach enables the replacement of irregular memory access operations with efficient computation.

\textcolor{black}{
The implementation of this mapping primarily consists of three stages. First, a vector addition unit subtracts the preset parameter $b$ from all input elements simultaneously, shifting the center of the data distribution close to zero. Then, a vector multiplication unit inverts the intermediate results to achieve the mathematical transformation $b - x$. For example, with $b = 123$, an input $x = 125$ is transformed to $123 - 125 = -2$, while an input $x = 122$ becomes $123 - 122 = 1$. This combined operation ensures that high-frequency values in the input data are mapped to smaller output values. Finally, we use shift operations to implement modulo $2^{n+1}$ and clear higher-bit information, restricting all output values to the predefined range $[0, 2^{n+1} - 1]$. Negative values $-c$ are converted to $2^{n+1} - c$ to achieve a wrapping effect. For instance, in the case where $n = 5$, the previous intermediate result of $-2$ is converted to $2^6 - 2 = 62$, while the positive result of $1$ remains unchanged. This approach ensures computational efficiency while maintaining injectivity, fully leveraging the vectorized computing capabilities of Ascend NPUs and avoiding inefficient branching operations.
}



\subsection{Intra-Segment Dependency Decoupled Scan 
}\label{sec:idd_scan}

The prefix sum (or scan) is a fundamental parallel primitive. However, its straightforward implementation on modern SIMD architectures, such as Ascend NPUs, is often hindered by stringent hardware constraints. The target operation is to compute the global prefix sum of all elements in an $N \times M$ tensor (where $M=16$), which is logically treated as a flattened 1D array. The primary bottleneck lies in the Ascend NPUs' memory model, particularly the 32-byte alignment requirement for operands in AscendC. Given that the data type used here is half-precision floating-point (\texttt{half}, 2 bytes), each row of $M$ elements in the tensor exactly occupies 32 bytes. Architectural constraints prohibit direct SIMD computation between elements residing within the same 32-byte memory segment. This effectively ``locks'' the direct intra-row prefix sum computation (e.g., $\text{row}[i] \mathrel{+}= \text{row}[i-1]$), as it requires operations between adjacent elements within a single hardware-indivisible block.

\begin{figure}[tbp]
    \centering
    \includegraphics[width=\linewidth]{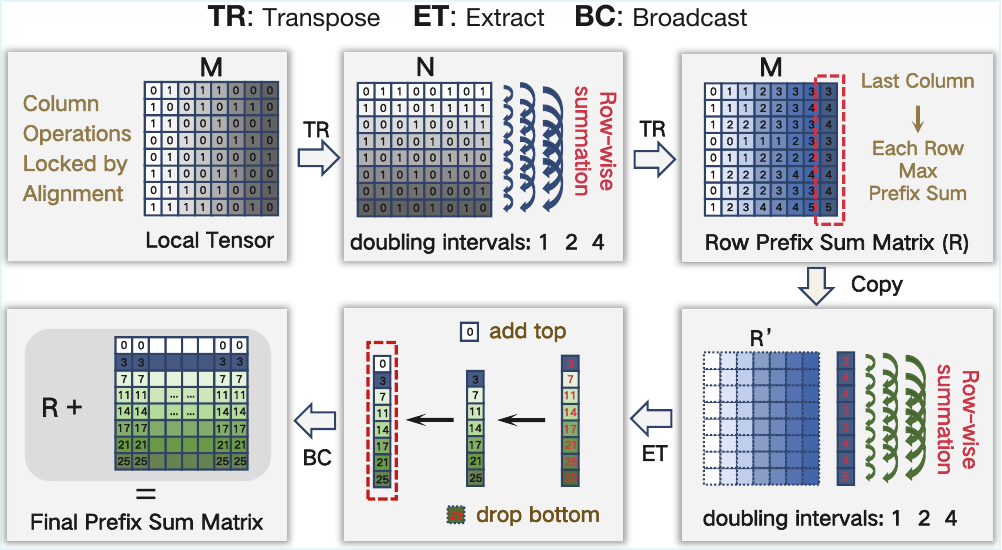}  
    \caption{\textcolor{black}{Prefix sum process. This figure illustrates the step-by-step implementation of the prefix sum algorithm, using an $8 \times 8$ binary local tensor $M$ as a working example.}}
    \label{fig:prefix_sum}
\end{figure}

To decouple this forbidden intra-segment dependency, we propose the \textit{Intra-Segment Dependency Decoupled Scan (IDD-Scan)}, a multi-stage algorithm that transforms the computation-bound problem into a series of hardware-friendly operations. The workflow of IDD-Scan is detailed below and illustrated in Figure~\ref{fig:prefix_sum}.

\textbf{Stage 1: Intra-Row Scan via Matrix Transposition.}
IDD-Scan first computes the prefix sum for each row by transforming it. The local $N \times M$ tensor is transposed into an $M \times N$ matrix. This distributes each original row's elements across $M$ new rows, converting intra-row dependencies into inter-row ones. A parallel prefix sum is then performed on each of the $N$ columns using vectorized additions in $\log_2(M)$ steps. The matrix is transposed back to its original $N \times M$ orientation, resulting in an intermediate matrix $\mathbf{R}$ where each row contains its correct local prefix sum.

\textbf{Stage 2: Inter-Row Propagation and Final Update.}
The intermediate matrix $\mathbf{R}$ contains row-local sums but lacks \textcolor{black}{exclusive cumulative offsets} from previous rows. A temporary copy $\mathbf{C}$ of $\mathbf{R}$ is created. A hierarchical scan is performed on $\mathbf{C}$'s rows in $\log_2(N)$ steps: at step $k$, each row $\mathbf{C}[i]$ adds values from row $\mathbf{C}[i - 2^k]$ element-wise. After completion, the last column of $\mathbf{C}$ holds the \textcolor{black}{inclusive cumulative offsets} for all rows. \textcolor{black}{Subsequently, the exclusive cumulative offsets are derived from the inclusive cumulative offsets by removing the bottom element and adding a zero element at the top.} This $N \times 1$ vector is broadcast across all columns to form an offset matrix. The final result is obtained by adding offset matrix element-wise to the saved matrix $\mathbf{R}$.

\begin{table*}[h]
\centering
\caption{Compression ratio results on different datasets.}
\label{tab:compression_ratio}
\resizebox{0.8\linewidth}{!}{
\renewcommand{\arraystretch}{1.1}
\begin{tabular}{l|l|ccccc|cc|ccc}
\toprule[0.4mm]
\multirow{2}{*}{\textbf{Arch}} & \multirow{2}{*}{\textbf{Compressors}} & \multicolumn{5}{c|}{\textbf{BF16}} & \multicolumn{2}{c|}{\textbf{FP16}} & \multicolumn{3}{c}{\textbf{FP32}} \\
\cmidrule(lr){3-7} \cmidrule(lr){8-9} \cmidrule(l){10-12}
& & Falcon & Qwen3-8B & DeepSeek & Qwen3-32B & Llama & Mistral & Diffusion & OLMo & BERT & Wav2Vec \\
\midrule[0.3mm]
CPU & ZipNN & 1.51 & 1.50 & 1.51 & 1.50 & 1.51 & 1.19 & 1.18 & 1.20 & 1.20 & 1.21 \\
\cmidrule{1-12}
\multirow{6}{*}{GPU} & NV\_Zstd & 1.28 & 1.27 & 1.28 & 1.27 & 1.29 & 1.09 & 1.08 & 1.08 & 1.08 & 1.08 \\
& NV\_Deflate & 1.28 & 1.27 & 1.28 & 1.27 & 1.29 & 1.09 & 1.08 & 1.08 & 1.08 & 1.08 \\
& NV\_GDeflate & 1.27 & 1.27 & 1.27 & 1.27 & 1.28 & 1.09 & 1.08 & 1.07 & 1.08 & 1.08 \\
& NV\_ANS & 1.25 & 1.24 & 1.25 & 1.25 & 1.27 & 1.08 & 1.04 & 1.06 & 1.04 & 1.03 \\
& NV\_Bitcomp & 1.33 & 1.32 & 1.33 & 1.32 & 1.32 & 1.13 & 1.07 & 1.14 & 1.14 & 1.15 \\         
& Diet\_ANS & 1.23 & 1.23 & 1.23 & 1.23 & 1.25 & 1.06 & 1.05 & 1.05 & 1.05 & 1.05 \\
& Diet\_Float & 1.48 & 1.47 & 1.48 & 1.47 & 1.48 & 1.17 & 1.16 & 1.19 & 1.19 & 1.19 \\
\cmidrule{1-12}
\multirow{2}{*}{NPU} & HANS & 1.34 & 1.34 & 1.35 & 1.34 & 1.33 & 1.09 & 1.05 & 1.14 & 1.13 & 1.13 \\ & \textbf{ENEC} & \textbf{1.36} & \textbf{1.36} & \textbf{1.37} & \textbf{1.35} & \textbf{1.36} & \textbf{1.12} & \textbf{1.09} & \textbf{1.15} & \textbf{1.15} & \textbf{1.15} \\
\bottomrule[0.4mm]
\end{tabular}
}
\vspace{-4mm}
\end{table*}

\subsection{Parameter Tuning}\label{ssec:design_hybrid}

This procedure analyzes exponent statistics to derive parameters that minimize the average compressed bit-length under a specialized cost model, following systematic phases:

\textbf{Phase 1: Statistical Pre-Analysis.} The initial phase involves constructing a histogram of the exponents extracted from the source data to obtain a frequency distribution for each unique exponent $x$. This distribution allows for the calculation of the probability $p(x)$ of each value occurring and the identification of the global minimum ($l$) and maximum ($h$) values in the exponents.

\textbf{Phase 2: Global Search for a Suitable Linear Mapping Parameter ($b$) and Base Bit-Width ($n$).} The algorithm performs an exhaustive search for a linear mapping parameter $b$ across its feasible integer domain. For each candidate $b$, the requisite base bit-width $n$ is computed as the minimal number of bits required to span the data range relative to $b$:

\begin{equation} \label{eq:bitwidth}
\footnotesize
n = \max\left( \lfloor \log_2(b - l) \rfloor + 1, \lceil \log_2(h - b) \rceil \right) + 1
\end{equation}

For each $(b, n)$ pair, all original exponents $x$ are transformed via a mapping:

\begin{equation} \label{eq:circular}
\footnotesize
y = (2^n - x + b) \% {2^n}
\end{equation}

A cost function $D$, defined as the probability-weighted sum of the transformed values, is evaluated:

\begin{equation} \label{eq:cost}
\footnotesize
D = \sum_x p(x) \cdot y
\end{equation}

The pair $(b^{*}, n^{*})$ that results in the minimum value for $D$ is selected. Using the selected parameters $(b^{*}, n^{*})$, the original exponents are transformed via Equation~\ref{eq:circular}. A statistical analysis of the bit-widths required for the transformed values $y$ is then performed. This yields the cumulative distribution function $p(m)$, representing the probability that a value $y$ can be represented using $m$ or fewer bits.

\textbf{Phase 3: Selection of Encoding Threshold ($m$) and Group Length ($L$).} In the final phase, the appropriate encoding threshold \(m\) and group length \(L\) are determined. Both \(L\) and \(m\) are treated as parameters. Since Ascend NPUs require data movement to be 32-byte aligned, we must set L $\ge$ $16$. The final parameter pair is selected through a joint search to minimize the expected bit length \(B_{\text{exp}}\):

\begin{equation} \label{eq:joint_selection}
\footnotesize
(m^{*}, L^{*}) = \arg \min_{m, L} \left[ B_{\text{exp}} = \frac{1}{L} + n + (m - n) \cdot p(m)^L \right]
\end{equation}


\textcolor{black}{
Under parameters $(m, L)$, expected bits per element $B_{\text{exp}}$ comprises amortized group mask overhead \(\frac{1}{L}\), base bit-width $n$, and threshold effect $(m-n) \cdot p(m)^L$. For such groups, each element uses $m$ instead of $n$ bits, saving $n-m$ per element, giving expected saving $(n-m) \cdot p(m)^L$ or equivalently adding $(m-n) \cdot p(m)^L$ to baseline $n$. Optimal \((m^*, L^*)\) minimize \(B_{\text{exp}}\) by balancing: larger \(L\) reduces \(\frac{1}{L}\) but lowers $p(m)^L$.
}

Upon completion of this process, the algorithm outputs the best set of selected parameters, denoted as the chosen linear mapping parameter $b^{*}$, the base bit-width $n^{*}$, the encoding threshold $m^{*}$, and the group length $L^{*}$.

\section{Experimental Evaluation}
\label{sec:evaluation}

\textbf{Platforms.}
\textcolor{black}{
All evaluations of NPU-based methods are performed on the Ascend NPU 910B2, which contains 24 Cube Units and 48 Vector Units (with a vector-to-cube unit ratio of 2:1). The host machine is powered by a HiSilicon Kunpeng-920 CPU running Ubuntu 22.04.5 LTS, with NPU driver version 25.0.rc1.1. Our algorithms are implemented in C++17 using the AscendC framework, with toolkit version 8.2.RC1.alpha002 and the same version for the operator kernel package. To ensure a fair comparison, CPU-based and GPU-based methods are evaluated on a performance-similar NVIDIA A800 GPU platform, which is equipped with Intel Xeon 8358P CPUs and uses CUDA version 12.6.
}

\begin{table}[h]
    \caption{Popular models' weights data used in this work.}
    \centering
    \resizebox{\linewidth}{!}{
    \renewcommand{\arraystretch}{1.1}
        \begin{tabular}{cccc}
            \toprule[0.5mm]
            \textbf{Type} & \textbf{Models} & \textbf{Size (GB)} & \textbf{Layer Size (MB)} \\
            \midrule[0.4mm]
            \multirow{5}{*}{BF16} & Falcon-7B  \cite{falcon} & 14.40 & 39.38{\small$\sim$}563.56 \\
             & Qwen3-8B  \cite{qwen3technicalreport} & 16.38 & 8.00{\small$\sim$}1187.00 \\
             & deepseek-llm-7b-base  \cite{deepseek-llm} & 13.82 & 32.00{\small$\sim$}800.00 \\
             & Qwen3-32B  \cite{qwen3technicalreport} & 65.60 & 10.00{\small$\sim$}1483.75 \\
             & Llama-3.1-8B-Instruct  \cite{dubey2024llama} & 16.06 & 8.00{\small$\sim$}1002.00 \\
            \midrule
            \multirow{2}{*}{FP16} & CapybaraHermes-2.5-Mistral-7B  \cite{Jiang2023Mistral7} & 14.50 & 8.00{\small$\sim$}250.02 \\
             & stable-video-diffusion-img2vid-fp16  \cite{blattmann2023stable} & 4.27 & $3.10 \times 10^{-5}${\small$\sim$}96.50 \\
            \midrule
            \multirow{3}{*}{FP32} & OLMo-1B-hf  \cite{Groeneveld2023OLMo} & 5.10 & 16.00{\small$\sim$}393.00 \\
             & bert-base-uncased  \cite{bert} & 0.40 & 0.01{\small$\sim$}89.42 \\
             & wav2vec2-large-xlsr-53-english  \cite{conneau2020unsupervised} & 1.20 & $4.88 \times 10^{-4}${\small$\sim$}32.00 \\
            \bottomrule[0.5mm]
        \end{tabular}
    }
    \label{tab:datasets_params}
\end{table}

\textbf{Datasets.} We conduct tests on the weights of ten popular open-source AI models, including three FP32 models, two FP16 models, and five BF16 models. All model weights are downloaded directly from the HuggingFace platform. It should be noted that current mainstream large models, such as Qwen and Llama, adopt the BF16 format; therefore, \textcolor{black}{BF16 is our primary focus, as analyzed in Section~\ref{sec:analysis}}. Detailed information about the datasets can be found in Table \ref{tab:datasets_params}.

\textbf{Compression Baselines.} 
We compare our approach against SOTA lossless compressors tailored for different hardware platforms: ZipNN~\cite{hershcovitch2025zipnn} for CPU, DietGPU~\cite{github-dietgpu} and nvCOMP~\cite{NVCOMPNV83:online} for GPU, and HANS~\cite{hans-doc} for Ascend NPUs. Since both DietGPU and nvCOMP include multiple compression algorithms, we use ``Diet'' to denote DietGPU and ``NV'' to denote nvCOMP (e.g., Diet\_ANS refers to the ANS algorithm in DietGPU). Our primary evaluation metrics are compression ratio, compression throughput, and decompression throughput. ZipNN and Diet\_Float employ tail exponent separation on model data, which serves as a key benchmark for achieving high compression ratios. In contrast, other methods highlight the efficiency advantages of our ENEC. 
\begin{figure*}[t]
    \centering
    \includegraphics[width=1.0\linewidth]{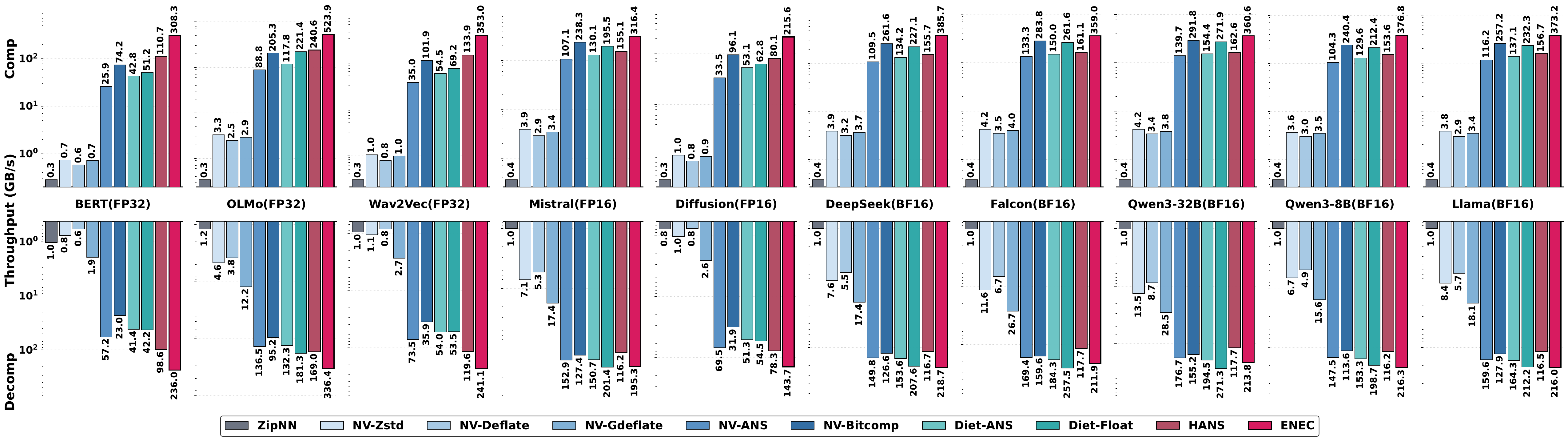}
    \caption{
    Throughput of compression (upper) and decompression (lower) across different datasets and methods. The Y-axis (Throughput) is on a logarithmic scale to visualize performance differences spanning several orders of magnitude.
    }
    \label{fig:comp_throughput}
    \vspace{-4mm} 
\end{figure*}

\textbf{Compression Configurations.} For the ENEC method, we perform offline automatic tuning using the search algorithm described in Section \ref{ssec:design_hybrid}, and adopt the resulting configurations (as shown in Table \ref{tab:best_parameters}) as the actual parameters for online compression in subsequent models. For all other methods, we use their default configurations. Specifically, for baseline methods such as nvCOMP and DietGPU, we utilize their default settings, which are designed to reflect the optimal trade-off between throughput and compression ratio. Although we also performed parameter tuning for these baselines, the resulting performance showed negligible difference. ENEC runs exclusively on the AIV units, which have significantly lower peak compute capacity than a full A800 GPU. Nevertheless, its competitive performance demonstrates the remarkable effectiveness of architectural specialization and justifies our throughput comparisons, highlighting the success of our hardware/software co-design for this dedicated task.

\textbf{End-to-End Inference Configurations.} We integrate ENEC into the HuggingFace Transformers inference framework and evaluate it on two mainstream large language models, Qwen3-32B and Falcon-40B. The evaluation metrics are \textcolor{black}{Time-To-First-Token} (TTFT) and \textcolor{black}{Time-Per-Output-Token} (TPOT), measured using fixed input and output lengths. During inference, decompression is performed layer-wise. We overlap the next layer’s decompression with the current layer’s forward. The baseline uses an uncompressed inference setup with partial CPU offloading, since a single Ascend 910B2 cannot hold the full model weights.


\subsection{Compression Ratio}\label{ssec:comp_ratio}
The compression ratio results for all evaluated models are summarized in Table~\ref{tab:compression_ratio}. Importantly, we have meticulously verified that all 10 models (across FP32, FP16, and BF16 precisions) listed in Table \ref{tab:datasets_params} achieve \textbf{bit-identical reconstruction} via our compression and decompression pipeline. 

For each model's weight data, we report the compression ratio by dividing the total size of all parameters before compression by the total size after compression. Since ZipNN, nvCOMP, DietGPU, and ENEC perform compression in file form, we first save the model's raw parameters as binary files before applying the respective file compression. For HANS, which only provides a Python API for compressing tensors, we load the model and then traverse each parameter tensor to conduct compression tests.


\textcolor{black}{
Table \ref{tab:compression_ratio} shows ENEC’s consistently strong performance across datasets. ENEC generally outperforms HANS and significantly exceeds nvCOMP’s general-purpose methods. This advantage stems from our specialized exponent-mantissa separation. Unlike general compressors that process floating-point values uniformly—diluting exponent frequency statistics—our targeted linear mapping enables more accurate modeling and higher compression ratios.
}

\textcolor{black}{
ENEC remains competitive against SOTA specialized compressors. While ZipNN (CPU) and DietGPU::Float (GPU) achieve higher ratios, ENEC’s performance is notably close on FP32 and FP16 datasets. This reflects a deliberate trade-off: ENEC prioritizes maximum throughput on resource-constrained Ascend NPUs via massive vectorization. While this architectural choice limits element-wise encoding flexibility—slightly reducing compression ratios—the substantial throughput gains meet high-performance data transmission demands. We argue this modest compromise is well justified by the significant practical efficiency.
}

\begin{figure*} [htbp]
  \includegraphics[width=1\linewidth]{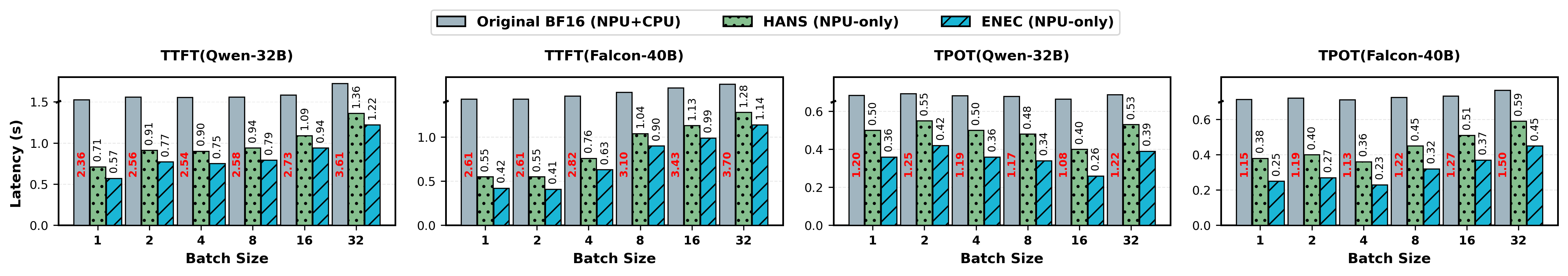}
  \caption{\textcolor{black}{End-to-end inference performance comparison of ENEC across different models and batch sizes. Latency metrics include Time-To-First-Token (TTFT) and Time-Per-Output-Token (TPOT). }}
  \label{fig:end2end}
  \vspace{-4mm}
\end{figure*}

\subsection{Compression and Decompression Throughput} \label{ssec:throughput}

In this section, we evaluate and compare ENEC's compression and decompression throughput against other lossless compressors. Throughput is calculated by dividing the raw data size by the total compression or decompression time. 

For ZipNN and DietGPU, we measure performance using their direct APIs. For nvCOMP, we use their benchmark code compiled with C++ and CUDA files, and report their direct presented results. For HANS and ENEC, we utilize the msprof profiling tool within the AscendC framework to assess kernel-level performance. This approach is adopted because HANS only provides a Python interface, and performance measurements at the Python level may differ significantly from the actual kernel-level performance. For both GPU and NPU platforms, reported throughput values measure device-side kernel execution time only, excluding PCIe transfer.

\textbf{Compression Throughput.} As shown in Figure ~\ref{fig:comp_throughput}, ENEC achieves significantly higher throughput across all data types compared to existing baseline platforms on CPUs, GPUs, and Ascend NPUs. Specifically, for BF16 models, ENEC achieves an average throughput of 372 GB/s, which is 987$\times$ higher than ZipNN—the SOTA compressor on CPUs—and 1.39$\times$ as much as nvCOMP:Bitcomp, the best-performing compressor on GPUs. It also outperforms HANS, another NPU-based compressor, by a factor of 1.36. For FP32 models, the performance gap remains consistent, with ENEC achieving throughput that is 1266$\times$ that of ZipNN, 3.08$\times$ that of Bitcomp, and 2.47$\times$ that of HANS. The performance for FP16 is similar to that of BF16, with ENEC achieving an average throughput of 263 GB/s, which is 767.33$\times$ faster than ZipNN, 0.58$\times$ higher than Bitcomp, and 2.27$\times$ that of HANS. These results strongly demonstrate the effectiveness of our lightweight and highly vectorized optimization approach tailored for Ascend NPUs' platforms.

Moreover, despite FP16’s 5-bit exponent limiting compression and adding design challenges, our ENEC still performs nearly on par with BF16, reaching up to 317 GB/s, showcasing its efficiency across data types. Furthermore, ENEC achieves up to 523 GB/s on FP32—nearly twice BF16’s speed. This is because we avoid compressing the mantissa. As data width doubles from 16-bit to 32-bit, the processed amount remains unchanged. Thus, compression time stays similar, significantly boosting throughput.

\textbf{Decompression Throughput.} \textcolor{black}{Decompression throughput is measured and presented using the same methodology as compression throughput. Results across various model datasets are also shown in Figure \ref{fig:comp_throughput}. Compared to compression, both HANS and ENEC exhibit lower decompression throughput. This is primarily due to operations like prefix sum computation and reverse gather, which are not well optimized on Ascend NPU architectures, increasing computational overhead.}

\textcolor{black}{Nevertheless, ENEC maintains a comprehensive throughput advantage; its decompression performance significantly surpasses all other algorithms, achieving speedups of up to 4.22$\times$ over NV-Bitcomp and 2.11$\times$ over HANS, except for a 6\% drop compared to Diet-Float on GPU for BF16 models. On FP16 and FP32 models, we still achieve an improvement of approximately 0.31–1.90 over Diet-Float. This gain stems from our optimizations in prefix sum computation and replacing lookup-table gather with linear mapping, which reduce memory access latency and enhance parallel efficiency.}

\begin{table}[htbp]
  \caption{ENEC compression parameters adopted by the majority of tensors in the models. }
  \centering
  \scriptsize
  \resizebox{\linewidth}{!}{
  \setlength{\tabcolsep}{5pt}            
  \renewcommand{\arraystretch}{1.1}     
  \renewcommand\tabularxcolumn[1]{m{#1}}
  \begin{tabularx}{\linewidth}{@{}%
      >{\centering\arraybackslash}m{6mm}  
      >{\centering\arraybackslash}X       
      >{\centering\arraybackslash}m{23mm} 
      >{\centering\arraybackslash}m{6mm}  
      >{\centering\arraybackslash}X       
      >{\centering\arraybackslash}m{23mm} 
      @{}}
    \toprule[0.3mm]
    \multicolumn{3}{c}{\textbf{BF16}} & \multicolumn{3}{c}{\textbf{FP16 \& FP32}} \\
    \cmidrule(lr){1-3}\cmidrule(lr){4-6}
    \textbf{Type} & \textbf{Models} & \textbf{(b, n, m, L)} &
    \textbf{Type} & \textbf{Models} & \textbf{(b, n, m, L)} \\
    \midrule[0.2mm]
    BF16 & Falcon     & \texttt{(122, 6, 3, 16)} & FP16 & Mistral   & \texttt{(7, 4, 3, 16)}  \\
    BF16 & Qwen3-8B   & \texttt{(123, 6, 3, 16)} & FP16 & Diffusion & \texttt{(11, 5, 3, 16)} \\
    BF16 & DeepSeek   & \texttt{(123, 6, 3, 16)} & FP32 & OLMo      & \texttt{(121, 6, 3, 16)} \\
    BF16 & Qwen3-32B  & \texttt{(122, 6, 3, 16)} & FP32 & BERT      & \texttt{(123, 6, 3, 16)} \\
    BF16 & Llama      & \texttt{(121, 6, 3, 16)} & FP32 & Wav2Vec   & \texttt{(125, 6, 3, 16)} \\
    \bottomrule[0.3mm]
  \end{tabularx}
  }
  \label{tab:best_parameters}
\end{table}

\subsection{End-to-End Inference Speedup}

We evaluate the baseline and the ENEC-integrated inference framework on two mainstream models. Because a single NPU cannot host the full model, the baseline keeps most weights on the NPU and offloads only essential parts to the CPU. Our inference evaluation includes 10 warm-up and 50 test runs, from which we compute the average TTFT and TPOT. \textcolor{black}{As shown in Figure \ref{fig:end2end}, ENEC consistently surpasses both the uncompressed baseline and HANS. On Qwen3-32B, ENEC reduces TTFT by up to 4.1× and TPOT by up to 3.9× compared to the baseline; relative to HANS, ENEC achieves up to 1.7× lower TTFT and 1.6× lower TPOT. On Falcon-40B, ENEC attains maximum reductions of 6.3× in TTFT and 4.9× in TPOT over the baseline, and up to 1.8× and 1.7× improvements over HANS, respectively.} These substantial improvements mainly result from eliminating the overhead of weight transfers and ENEC's superior decompression throughput. For instance, with a batch size of 4, transferring certain Falcon layer weights from CPU to NPU accounts for nearly 80\% of execution time.

\subsection{Evaluation of Parameter Tuning} \label{ssec:param_tune}

This section presents a detailed experimental analysis of several parameters in our compressor, supplementing the theoretical analysis in Section~\ref{ssec:design_hybrid}. We investigate the impact of data block size, group length ($L$), quantization parameters ($n$ and $m$), and the integer transformation parameter ($b$).

\begin{figure}[htbp]
    \centering
    \includegraphics[width=.9\linewidth]{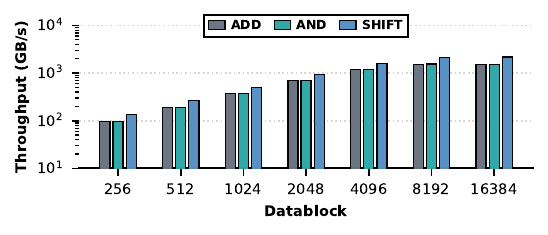}
    \caption{Performance of several common operations under fixed input conditions across different data block sizes.}
    \label{fig:operation_throughput}
\end{figure}

\textcolor{black}{
\textbf{Data Block Size:} This parameter primarily dictates system throughput with negligible compression impact. Throughput scales with block size (Fig.~\ref{fig:operation_throughput}). For performance, we choose a 16,384-element block, avoiding 32,768 as its memory footprint exceeds UB's 192KB limit on Ascend NPUs.
}

\textcolor{black}{
\textbf{Group Length ($L$):} Our experimental results, summarized in Table~\ref{tab:best_parameters}, demonstrate that a group length of $L=16$ consistently yields the optimal balance between granularity and compression efficiency. Notably, this value remains stable across different data types (FP32, BF16, and FP16), suggesting that $L=16$ provides an ideal alignment with the NPU's vectorized memory access patterns.
}

\textcolor{black}{
\textbf{Quantization Parameters ($b, n, m$):} These parameters are determined via the joint search optimization algorithm (Section~\ref{ssec:design_hybrid}). We observe that for FP32 and BF16, the quantization parameters $n$ and $m$ remain highly consistent, with only minor variations in $b$ due to data distribution differences. For FP16, despite its narrower 5-bit exponent, the optimal value for $n$ remains 6, matching the other formats.
}

\subsection{\textcolor{black}{Parameter Robustness and Transferability}}

\textcolor{black}{
To assess ENEC robustness, we perform cross-model sensitivity analysis applying optimal parameters from DeepSeek-V3 to other LLMs (e.g., Qwen-3) without re-tuning.
}

\begin{table}[htbp]
    \caption{\textcolor{black}{Compression performance on other models using the parameters searched from DeepSeek-V3. ``Optimal'' is the performance achieved using the parameters searched on the target model.}}
    \centering
    \resizebox{\linewidth}{!}{
    \renewcommand{\arraystretch}{1.1}
        \begin{tabular}{ccc}
            \toprule[0.5mm]
            \textbf{Models} & \textbf{Compression Ratio} & \textbf{Comp. / Decomp. Thr. (GB/s)} \\
            \midrule[0.4mm]
            Falcon-7B \cite{falcon} & 1.34 (Optimal: 1.34, 0\%↓) & 367 / 213 (Optimal: 359 / 212) \\
            Qwen3-8B \cite{qwen3technicalreport} & 1.35 (Optimal: 1.35, 0\%↓) & 375.2 / 220.1 (Optimal: 377 / 216) \\
            Qwen3-32B \cite{qwen3technicalreport} & 1.34 (Optimal: 1.34, 0\%↓) & 351 / 209 (Optimal: 361 / 214) \\
            Llama-3.1-8B-Instruct \cite{dubey2024llama} & 1.31 (Optimal: 1.36, 5\%↓) & 338 / 199 (Optimal: 373 / 216) \\
            \bottomrule[0.5mm]
        \end{tabular}
    }
    \label{tab:Transferability}
\end{table}

\textcolor{black}{
Results in Table \ref{tab:Transferability} show that the optimal parameters searched from DeepSeek-V3 transfer well across multiple models. For Falcon-7B, Qwen3-8B, and Qwen3-32B, compression remains fully lossless, and throughput even improves slightly, confirming ENEC stably delivers high performance under fixed structural parameters (group length $L$ and bit-packing logic $n$, $m$). For Llama-3.1-8B-Instruct, a minor 5\% loss in compression ratio and slight throughput decrease occur, but absolute performance remains relatively high. These fluctuations likely stem from differences in weight distribution across models, not algorithm design issues. Overall, the parameters searched by DeepSeek-V3 achieve zero-loss migration on most models, validating their effectiveness and robustness across different architectures.
}

\subsection{Ablation Study}
\label{ssec:ablation}
\textcolor{black}{
We perform an ablation study starting from baseline ENEC V0 (Section \ref{sec:basic}). V1 adds bit-width quantization with hierarchical halving bit-packing (Section \ref{sec:bit_packing}); V2 introduces vectorized branch-free integer transformation (Section \ref{sec:branch_free}); and V3 incorporates \textit{IDD-Scan} for decompression (Section \ref{sec:idd_scan}). We evaluate all versions and block size impact on DeepSeek~\cite{deepseek-llm} using ratio and throughput.
}

\begin{table}[htbp]
\caption{Compression ratios with different \textcolor{black}{input file} sizes (MB).}
\label{tab:ratio_bs}
\centering
\resizebox{\linewidth}{!}{
        \renewcommand{\arraystretch}{1.1}
        \begin{tabular}{l l c c c c c c c c c c}
        \toprule[0.5mm]
        \textbf{Arch} & \textbf{Compressors} & 1 & 2 & 4 & 8 & 16 & 32 & 64 & 128 & 256 & 512 \\
        \midrule[0.4mm]
        CPU & ZipNN & 1.51 & 1.51 & 1.51 & 1.51 & 1.51 & 1.51 & 1.51 & 1.50 & 1.50 & 1.50 \\
        \cmidrule(r){1-12}
        \multirow{7}{*}{GPU} & NV\_Zstd & 1.28 & 1.28 & 1.28 & 1.28 & 1.28 & 1.28 & 1.28 & 1.27 & 1.27 & 1.28 \\
         & NV\_Deflate & 1.28 & 1.28 & 1.28 & 1.28 & 1.28 & 1.28 & 1.28 & 1.27 & 1.27 & 1.28 \\
         & NV\_GDeflate & 1.27 & 1.27 & 1.27 & 1.27 & 1.27 & 1.27 & 1.27 & 1.27 & 1.27 & 1.27 \\
         & NV\_ANS & 1.20 & 1.20 & 1.20 & 1.20 & 1.20 & 1.23 & 1.25 & 1.26 & 1.26 & 1.26 \\
         & NV\_Bitcomp & 1.33 & 1.33 & 1.33 & 1.33 & 1.33 & 1.33 & 1.33 & 1.36 & 1.36 & 1.36 \\
         & Diet\_ANS & 1.23 & 1.23 & 1.23 & 1.23 & 1.23 & 1.23 & 1.23 & 1.23 & 1.23 & 1.23 \\
         & Diet\_Float & 1.48 & 1.48 & 1.48 & 1.48 & 1.48 & 1.48 & 1.48 & 1.48 & 1.48 & 1.48 \\
        \cmidrule(r){1-12}
        \multirow{2}{*}{NPU} & HANS & 1.30 & 1.30 & 1.25 & 1.32 & 1.33 & 1.34 & 1.35 & 1.35 & 1.35 & 1.35 \\
         & \textbf{ENEC} & \textbf{1.37} & \textbf{1.37} & \textbf{1.37} & \textbf{1.37} & \textbf{1.38} & \textbf{1.37} & \textbf{1.38} & \textbf{1.37} & \textbf{1.37} & \textbf{1.38} \\
        \bottomrule[0.5mm]
        \end{tabular}
}
\end{table}

\begin{figure}[htbp]
    \centering
    \includegraphics[width=\linewidth]{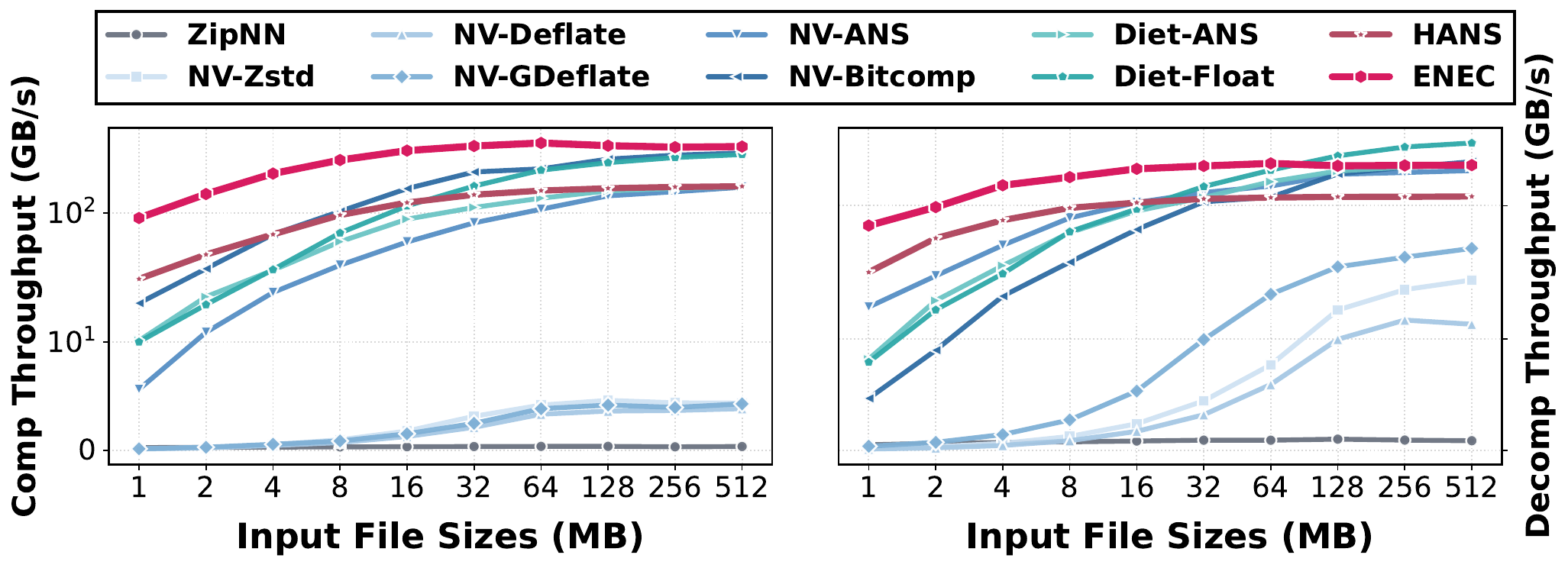}
    \caption{\textcolor{black}{Comparison of throughput performance across various methods under different input file sizes.}}
    \label{fig:datablock_throughput}
\end{figure}

\textcolor{black}{\textbf{Analysis of Different Input File Sizes.}
Using DeepSeek as a case study, we analyzed compression across varying input sizes by partitioning parameters into different segment sizes and compressing them independently. As shown in Table \ref{tab:ratio_bs} and Figure \ref{fig:datablock_throughput}, ENEC consistently outperforms baselines across all sizes, maintaining a stable advantage.}

\begin{figure}[ht]
    \centering
    \includegraphics[width=0.9\linewidth]{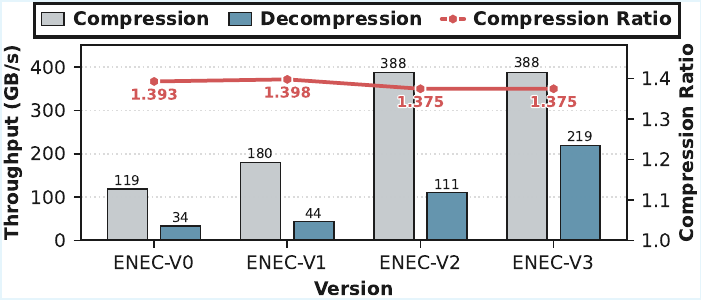}
    \caption{Performance of several ENEC versions.}
    \label{fig:version_throughput}
\end{figure}

\textbf{Analysis \textcolor{black}{of} Different ENEC Versions.}
\textcolor{black}{
Figure \ref{fig:version_throughput} compares compression ratio and throughput across versions. V0 achieves a high ratio via frequency-based statistical mapping. V1 slightly improves on V0 by reducing per-group metadata from 4 bits to 1 bit. V2 and V3 adopt branchless integer transform; its linear mapping approximates the frequency distribution, slightly reducing the ratio. For throughput, V1 improves compression and decompression by nearly 30\% over V0 using bit-width quantization with hierarchical bit-packing, replacing multiplication/division. V2 replaces slow gather with vectorized branch-free integer transformation, nearly doubling throughput on vector units. V3 further optimizes decoding with \textit{IDD-Scan}, boosting decompression throughput by nearly 100\% over V2.
}

\section{\textcolor{black}{Discussion}}

\textcolor{black}{
This section discusses ENEC's generality and its adaptability to CPUs and GPUs, highlighting its modular design, which simplifies porting and optimization. We then derive architectural implications from our software–hardware co-design, offering insights for future accelerators to better support efficient lossless compression.
}

\subsection{\textcolor{black}{Generality of Proposed Design}}
\label{ssec:Generality}

\textcolor{black}{
Although ENEC's low-level implementation is deeply customized for the specific constraints of Ascend NPUs, its core design philosophy and algorithmic framework possess broad generality. The framework can be readily adapted to other accelerators, including NVIDIA GPUs.
}

\begin{table}[htbp]
    \caption{\textcolor{black}{Performance comparison of ENEC and its implementations on other platforms using Qwen-32B.}}
    \centering
    \resizebox{.82\linewidth}{!}{
    \renewcommand{\arraystretch}{1.1}
        \begin{tabular}{cccc}
            \toprule[0.5mm]
            \textbf{Hardware} & \textbf{Compressors} & \textbf{CR} & \textbf{Comp. / Decomp. Thr. (GB/s)} \\
            \midrule[0.4mm]
            \multirow{2}{*}{CPU} & ZipNN & 1.50 & 0.4 / 1.0\\
             & \textbf{ENEC-CPU} & \textbf{1.35} & \textbf{3.91 / 1.56} \\
             \midrule
            \multirow{5}{*}{GPU} & NV-ANS & 1.24 & 139.7 / 176.7 \\
             & NV-Bitcomp & 1.32 & 291.8 / 155.2 \\
             & Diet-ANS & 1.23 & 154.4 / 194.5 \\
             & Diet-Float & 1.47 & 271.9 / 271.3 \\
             & \textbf{ENEC-GPU-V0} & \textbf{1.34} & \textbf{291.3 / 269.4} \\ 
             & \textbf{ENEC-GPU-V1} & \textbf{1.35} & \textbf{419.2 / 421.0} \\
            \midrule
            \multirow{2}{*}{NPU} & HANS & 1.34 & 162.6 / 117.7\\
             & \textbf{ENEC} & \textbf{1.35} & \textbf{360.6 / 213.8} \\
            \bottomrule[0.5mm]
        \end{tabular}
    }
    \label{tab:Generality}
\end{table}

\textcolor{black}{
\textbf{Cross-Architecture Portability.} The core components of ENEC—bit-width quantization, branch-free transformation, and hierarchical halving
bit-packing—are designed to maximize efficiency by converting complex conditional logic into regularized, vectorized bitwise operations. These design principles are inherently portable across parallel computing architectures. To validate its portability, we implemented a baseline version (ENEC-GPU-V0) on the A800 GPU that strictly follows the original execution flow. As shown in Table~\ref{tab:Generality}, ENEC-GPU-V0 achieves a compression ratio of 1.34× and throughput of 291.3/269.4 GB/s, which is comparable to the NPU version's 1.35× ratio and 360.6/213.8 GB/s throughput. This demonstrates that ENEC's fundamental logic is independent of hardware instruction sets and can be seamlessly migrated across diverse architectures.
}

\textcolor{black}{
\textbf{Hardware-Specific Optimizations.} ENEC's modular design enables deep hardware-tailored tuning by substituting computational primitives. For instance, hierarchical halving bit-packing can be optimized using platform-specific shuffle or shift instructions. While the NPU version uses IDD-Scan to bypass alignment constraints, the GPU version (ENEC-GPU-V1) invokes optimized parallel prefix-sum from the CUB library, leverages warp-level intrinsics for fast communication, and utilizes vectorized memory access. These optimizations yield a 1.56× throughput improvement over the baseline (ENEC-GPU-V0). Similarly, our CPU implementation employs AVX2 for BF16 transformations and BMI2 PEXT for non-byte-aligned compression, achieving 1.56× the throughput of ZipNN in a single-core configuration.}

\subsection{\textcolor{black}{Architectural Implications for Future Accelerators}}
\label{ssec:Architectural}

\textcolor{black}{
The ENEC co-design highlights key principles for future architectures with compression acceleration:
\begin{itemize}[topsep=2pt,leftmargin=0.2em]
\item \textbf{Branch-free execution support:} Architectures should natively support branch-free execution via bit-manipulation and predication, aligning with prior compression studies \cite{shah2023lightweight}.
\item \textbf{Modular operator libraries:} Standardized interfaces for modular operator libraries enable vendor-optimized primitives (e.g., scans, gathers) while preserving portability.
\item \textbf{Memory subsystem co-design:} Memory subsystems should be co-designed with data layouts to reduce the overhead of non-contiguous access.
\item \textbf{Compression-specific hardware support:} We advocate for lightweight variable-length decoding near vector registers, dedicated per-lane bit-extraction, and fast parallel prefix-sum units to replace multi-stage workarounds such as IDD-Scan.
\item \textbf{Low-precision datapath specialization:} Specialized datapaths for low-precision formats should decouple exponent and mantissa processing and support hardware-level adaptive bit-width packing.
\end{itemize}
}

\textcolor{black}{
While some insights (e.g., the need for branch-free execution) echo prior work, this study is the first to systematically articulate them in the context of accelerator design.
}

\section{Related Work}
\label{sec:related}

The most relevant works to ENEC are DietGPU~\cite{github-dietgpu}, an efficient lossless compressor for GPUs, and HANS~\cite{hans-doc}, which is designed for Ascend NPUs. DietGPU, developed by Facebook Research, provides fast and specialized lossless compression on NVIDIA GPUs for both machine learning and HPC workloads. It includes a GPU-optimized ANS entropy coder to achieve high throughput and exposes two interfaces: a general ``ANS'' mode and a specialized ``Float'' mode tailored for model data. The ``Float'' mode compresses floating-point values by isolating and encoding only their exponents, a common strategy in weight compression. HANS~\cite{hans-doc} is an efficient closed-source lossless compressor developed by Huawei on Ascend NPUs.

\section{\textcolor{black}{Conclusion and Future Work}}
\label{sec:conclusion}


This paper presents a lossless compressor tailored for Ascend NPUs, specifically for model weight compression. Experiments show that ENEC achieves compression throughput of 263–523 GB/s and decompression throughput of 188–336 GB/s on FP32, FP16, and BF16 model weights. On 910B2 NPU, it delivers 2.47× higher compression throughput and 2.11× faster decompression throughput than SOTA methods, while also achieving superior compression ratios. It further demonstrates excellent performance in end-to-end inference scenarios. \textcolor{black}{Efficient compression is achieved through hardware-informed decomposition and data-dependent pipelining, balancing efficiency and compression ratio via an ``approximate majority + precise correction'' paradigm with trade-offs centered on system bottlenecks. As hardware and architecture technology evolves, we plan to extend support and performance optimizations to a broader range of low-precision and integer data formats.}

\section*{Acknowledgments}
\small
This work was supported by the Innovation Funding of ICT, CAS (Grant No. E461050), the National Key Research and Development Program of China (Grant No. 2025YFB3003702), and the National Natural Science Foundation of China (Grant Nos. 62032023 and T2125013). The AI-driven experiments, simulations and model training were performed on the robotic AI-Scientist platform of Chinese Academy of Sciences.

\bibliographystyle{IEEEtranS}
\bibliography{refs}

\newpage

\appendix
\section{Artifact Appendix}

\subsection{Abstract}

The source code of ENEC is available at \url{https://github.com/jinwuyang/ENEC_ISCA_AE}. The NPU kernel implementation can be found in the \textbf{csrc/} directory and the Python-based data processing, parameter search, and profiling scripts can be found in the \textbf{python/} directory. Since this paper includes a large number of experiments that in aggregate will take 24 hours to fully test all model compression tasks, the instructions here will focus on reproducing the results for \textbf{Qwen3-32B}. The workflow to reproduce other models is very similar.

\subsection{Artifact check-list (meta-information)}

{\small
\begin{itemize}
  \item \textbf{Algorithm: } ENEC ( Efficient NPU-Enhanced Compression).
  \item \textbf{Program: } Python 3.9, C++/Ascend C.
  \item \textbf{Compilation: } CMake 3.10+, GCC 7.5+ (aarch64), Ascend CANN Toolkit.
  \item \textbf{Transformations: } Layer-wise weight splitting.
  \item \textbf{Binary: } Compiled NPU operators (.so files) and executables.
  \item \textbf{Data set: } Qwen3-32B.
  \item \textbf{Run-time environment: } \url{Ubuntu 22.04}, \url{CANN 8.2.RC1.alpha002}.
  \item \textbf{Hardware: } Ascend 910B2 NPU.
  \item \textbf{Run-time state: } Isolated NPU execution.
  \item \textbf{Execution: } Automated profiling via msprof.
  \item \textbf{Metrics: } Compression Ratio (CR), Throughput (GB/s).
  \item \textbf{Output: } CSV reports and summary files.
  \item \textbf{Experiments: } Data preparation, parameter search, performance benchmarking, and inference.
  \item \textbf{How much disk space required (approximately)?: } 200 GB.
  \item \textbf{How much time is needed to prepare workflow (approximately)?: } 1 hours.
  \item \textbf{How much time is needed to complete experiments (approximately)?: } 3 hours.
  \item \textbf{Publicly available?: } Yes.
  \item \textbf{Code licenses: } BSD-3.
\end{itemize}
}

\subsection{Description}

\subsubsection{How to access}
The source code is available at \url{https://github.com/jinwuyang/ENEC_ISCA_AE}. The repository is organized into \textbf{csrc/} (NPU kernels), \textbf{python/} (test tools).

\subsubsection{Hardware dependencies}
The artifact requires an \textbf{Ascend 910B2 NPU} platform with \textbf{aarch64} architecture. 


\subsubsection{Software dependencies}
\begin{itemize}
    \item \textbf{CANN Stack:} Ascend-CANN-toolkit and Kernels 8.2.RC1.alpha002.
    \item \textbf{Python Libraries:} torch 2.5.1, torch\_npu 2.5.1.post3, and standard data science stack (numpy, pandas, scipy).
    \item \textbf{ATB Library:} Recommended version 8.0.0.
\end{itemize}

\subsubsection{Data sets}

The evaluation of ENEC encompasses a diverse set of model weights, categorized by their data precision formats. By default, the data\_prepare.sh script only downloads \textbf{Qwen3-32B} to minimize preparation time and disk usage. However, the data\_prepare.sh script provides commented options to download all other models listed below (e.g., DeepSeek-LLM-7B-Base, Falcon-7B, etc.) for users who wish to reproduce the complete set of experiments.

\noindent\textbf{BF16 Models:} 
\begin{lstlisting}[language=bash, xleftmargin=0.5em, xrightmargin=0.5em]
DeepSeek-LLM-7B-Base
Falcon-7B
Qwen3-8B
Llama-3.1-8B-Instruct
Qwen3-32B
Falcon-40B
\end{lstlisting}
\noindent\textbf{FP16 Models:}  
\begin{lstlisting}[language=bash, xleftmargin=0.5em, xrightmargin=0.5em]
CapybaraHermes-2.5-Mistral-7B
stable-video-diffusion-img2vid-fp16
\end{lstlisting}
\noindent\textbf{FP32 Models:} 
\begin{lstlisting}[language=bash, xleftmargin=0.5em, xrightmargin=0.5em]
OLMo-1B-hf
bert-base-uncased
wav2vec2-large-xlsr-53-english
\end{lstlisting}

\subsection{Installation}
\subsubsection{Install CANN Toolkit and Kernels}
Download the following files from \url{https://www.hiascend.com/developer/download/community/result?module=cann&cann=8.2.RC1.alpha002}:
    \begin{itemize}
        \item \url{Ascend-cann-toolkit_8.2.RC1.alpha002_linux-aarch64.run}
        \item \url{Ascend-cann-kernels-910b_8.2.RC1.alpha002_linux-aarch64.run}
    \end{itemize}
    Then run the following commands:
\begin{lstlisting}[language=bash, xleftmargin=0.5em, xrightmargin=0.5em]
# Add executable permissions
chmod +x Ascend-cann-toolkit_8.2.RC1.alpha002_linux-\aarch64.run
chmod +x Ascend-cann-kernels-910b_8.2.RC1.alpha002_linux-\aarch64.run
# Verify the installers
./Ascend-cann-toolkit_8.2.RC1.alpha002_linux-aarch64.run --\check
./Ascend-cann-kernels-910b_8.2.RC1.alpha002_linux-aarch64.\run --\check
# Install
./Ascend-cann-toolkit_8.2.RC1.alpha002_linux-aarch64.run --\install --install-path=/your/path
./Ascend-cann-kernels-910b_8.2.RC1.alpha002_linux-aarch64.\run --install --install-path=/your/path
source /your/path/ascend-toolkit/set_env.sh
\end{lstlisting}

\subsubsection{Configure the Conda environment}
Create a Python 3.9 environment and install NPU-specific PyTorch and dependencies:
\begin{lstlisting}[language=bash, xleftmargin=0.5em, xrightmargin=0.5em]
conda create -n enec python=3.9 -y
conda activate enec
pip install pandas numpy==1.24.3 transformers==4.30.0 jinja2 \decorator attrs psutil absl-py cloudpickle ml-dtypes scipy \tornado pyyaml
wget https://download.pytorch.org/whl/cpu/torch-2.5.1-cp39-\cp39-manylinux_2_17_aarch64.manylinux2014_aarch64.\whl 
pip install torch-2.5.1-cp39-cp39-manylinux_2_17_aarch64.\manylinux2014_aarch64.whl
wget https://gitee.com/ascend/pytorch/releases/download/v7.1.0.2-pytorch2.5.1/torch_npu-2.5.1.post3-cp39-cp39-\manylinux_2_17_aarch64.manylinux2014_aarch64.whl
pip install torch_npu-2.5.1.post3-cp39-cp39-manylinux_2_17_\aarch64.manylinux2014_aarch64.whl
\end{lstlisting}

\subsubsection{Verify the environment}
Run a simple NPU tensor operation to confirm correct setup:
\begin{lstlisting}[language=bash, xleftmargin=0.5em, xrightmargin=0.5em]
python3 -c "import torch; import torch_npu; a = torch.randn(3, 4).npu(); print(a + a)"
\end{lstlisting}
If the output is normal, the environment is normal.

\subsubsection{Build} Clone the repository and run build\_csrc.sh (1 hour).
\begin{lstlisting}[language=bash, xleftmargin=0.5em, xrightmargin=0.5em]
git clone https://github.com/jinwuyang/ENEC_ISCA_AE.git
chmod 777 -R ENEC_ISCA_AE
cd ENEC_ISCA_AE
bash build_csrc.sh
\end{lstlisting}


\subsection{Experiment workflow}
\subsubsection{Data Preparation} 

 Execute data\_prepare.sh to download and split the model weights. By default, the script only downloads and processes \textbf{Qwen3-32B} (1 hour). To test other models (e.g., DeepSeek-LLM-7B, Falcon-40B), simply uncomment the corresponding lines in data\_prepare.sh.
\begin{lstlisting}[language=bash, xleftmargin=0.5em, xrightmargin=0.5em]
bash data_prepare.sh
\end{lstlisting}

\subsubsection{Performance Testing} Run compressor\_test.sh to measure the compression ratio and throughput. This script automates parameter searching, compression/decompression profiling, and global analysis. At the end of the execution, it also outputs the end-to-end inference results (2 hours).
\begin{lstlisting}[language=bash, xleftmargin=0.5em, xrightmargin=0.5em]
source /your/path/ascend-toolkit/set_env.sh
bash compressor_test.sh
\end{lstlisting}


\subsection{Evaluation and expected results}

\subsubsection{Optimal parameter search results}
The following results show the expected outputs for the Qwen3-32B model:
\begin{lstlisting}[language=bash, xleftmargin=0.5em, xrightmargin=0.5em]
              BF16 Model Compression Results              
========================================
File Processed:      hyperparams_results.csv
Total Elements:      32,761,446,400
-------------------------------------------------
Original BF16 Size:     62487.50 MB
ENEC Compressed Size:    46265.99 MB
Compression Ratio (CR):      1.35x
Model Avg Bit-width:      11.8465 bits/element
Exponent Avg Bit-width:    3.8465 bits/element
Formula Avg CR*:               1.35 x
\end{lstlisting}
The optimal parameter search results are organized within the \textbf{param\_search\_enec/} directory. Each model subfolder (e.g., BF16/Qwen3-32B) provides:
\begin{itemize}
    \item \textbf{hyperparams\_results.csv}: An exhaustive list of optimal parameters for every model tensor.
    \item \textbf{param\_combinations\_stats.txt}: A comprehensive statistical report of the search results.
\end{itemize}

\subsubsection{Compression Ratio and Throughput}
The file summary\_enec.csv summarizes the compression ratio, compression throughput, and decompression throughput of ENEC on \textbf{11 models}, corresponding to Table~\ref{tab:compression_ratio} and Figure~\ref{fig:comp_throughput} in the paper. The expected results for these 11 models are presented as follows:

\begin{lstlisting}[language=bash, xleftmargin=0.5em, xrightmargin=0.5em]
--- Summary Data Preview ---
model_name dtype  compression_ratio_CR  compress_throughput_GBps  decompress_throughput_GBps
Llama-3.1-8B-Instruct  BF16                  1.36                     376.8                       219.4
Qwen3-32B  BF16                  1.35                     366.3                       217.1
Qwen3-8B  BF16                  1.36                     388.1                       222.7
deepseek-llm-7b-base  BF16                  1.37                     391.2                       223.1
falcon-40b  BF16                  1.37                     369.1                       217.2
falcon-7b  BF16                  1.36                     364.6                       215.8
CapybaraHermes-2.5-Mistral-7B  FP16                  1.12                     317.0                       195.5
stable-video-diffusion-img2vid  FP16                  1.09                     223.4                       148.0
OLMo-1B-hf  FP32                  1.15                     538.6                       348.7
bert-base-uncased  FP32                  1.15                     329.1                       252.8
wav2vec2-large-xlsr-53-english  FP32                  1.15                     372.1                       254.6
\end{lstlisting}

\subsubsection{End-to-End Inference Latency}
Figure~\ref{fig:end2end} in the paper shows the end-to-end inference latency and speedup over the baseline (uncompressed with CPU offloading) for both Qwen3-32B and Falcon-40B under different batch sizes. For brevity, we only present the results for \textbf{Qwen3-32B} with \textbf{batch size = 1}. The expected results are presented as follows:

\begin{lstlisting}[language=bash, xleftmargin=0.5em, xrightmargin=0.5em]
[Inference: Qwen3-32B]
  Configuration: size=61.02 GB, throughput=217.05 GB/s
  baseline TTFT: 2.36064 s
  baseline TPOT: 1.1951 s
  ENEC TTFT: 0.556342 s (Speedup: 4.24x)
  ENEC TPOT: 0.361142 s (Speedup: 3.31x)
  Result saved to: Latency_Qwen3-32B_BF16.csv
\end{lstlisting}

\end{document}